\documentclass[aps,prd,twocolumn,amsmath,amssymb,groupedaddress,nofootinbib]{revtex4-2}

\usepackage{graphicx}
\usepackage{dcolumn}
\usepackage{bm}
\usepackage{multirow} 

\begin{document}

\preprint{APS/123-QED}

\title{Tilt-to-length noise subtraction with pointing jitters from closed-loop dynamics for TianQin}

\def\addressa{MOE Key Laboratory of TianQin Mission, TianQin Research Center for Gravitational Physics $\&$ School of Physics and Astronomy, Frontiers Science Center for TianQin, Gravitational Wave Research Center of CNSA, Sun Yat-sen University (Zhuhai Campus), Zhuhai 519082, China.}

\author{Yuzhou Fang}
\altaffiliation{Joint first authors}
\affiliation{\addressa}

\author{Dexuan Zhang}
\altaffiliation{Joint first authors}
\affiliation{\addressa}

\author{Dezhi Wang}
\affiliation{\addressa}

\author{Xuefeng Zhang}
\email[]{zhangxf38@sysu.edu.cn}
\affiliation{\addressa}

\author{Huizong Duan}
\email[]{duanhz3@sysu.edu.cn}
\affiliation{\addressa}

\author{Hongyin Li}
\affiliation{\addressa}

\author{Junxiang Lian}
\affiliation{\addressa}

\author{Guoying Zhao}
\email[]{zhaoguoying@sysu.edu.cn}
\affiliation{\addressa}

\date{\today}

\begin{abstract}
TianQin is a proposed space-based mission for gravitational wave detection, employing a constellation of three drag-free satellites in high Earth orbits to form a laser interferometric observatory. A critical technical challenge is mitigating tilt-to-length (TTL) coupling noise, which is expected to be the third dominant noise source after laser frequency and clock noises. This noise is unavoidable in the presence of the residual angular movement of satellites, movable optical subassemblies (MOSAs), and test masses (TMs), and needs to be subtracted after reducing the first two types of noises using time-delay interferometry (TDI). Previous works have shown that TTL coupling coefficients can be estimated from the null TDI channel $\zeta$ and used for noise subtraction in other combinations. However, it was found that correlated MOSA yaw jitters have a negative impact on the TTL calibration, and the effects of realistic residual angular jitters from drag-free and pointing control (DFPC) are yet to be investigated. In this paper, we use closed-loop DFPC simulations to generate more realistic jitters in the science mode and test TTL calibration capability. Our simulations reveal that rotating only one MOSA is more favorable, compared to symmetrically rotating two MOSAs, for enhancing the accuracy of TTL coefficient estimation, while employing only high-frequency data (0.1 - 1 Hz). Moreover, we propose two other methods to further improve estimation accuracy. Firstly, using different null channel combinations, such as $C_3^{14}$, enhances the least squares estimation accuracy even in the case of high correlations in MOSAs' yaw jitters. Secondly, injecting different sinusoidal artificial maneuvers to the six MOSAs also shows improvements. These methods can help TianQin to meet the 0.3 pm/Hz$^{1/2}$ requirement after the TTL noise subtraction. 
\end{abstract}

\maketitle


\section {Introduction}\label{sec: intro}
TianQin is a planned space observatory set for launch in the 2035s, targeting the detection of gravitational waves (GWs) in the 0.1 mHz to 1 Hz frequency band~\cite{Luo2016, luo2025progress}. Arranged in an almost equilateral triangle formation with three identical satellites, TianQin demands highly accurate displacement measurements between two test masses (TMs) at the picometer scale. TianQin's arm lengths are approximately 1.73 $\times10^5$ km, which significantly exposes the detector to laser frequency noise. However, by applying the time-delay interferometry (TDI) algorithm~\cite{tinto2021time}, this predominant laser frequency noise can be effectively mitigated during postprocessing.

Besides the noise from the laser frequency, tilt-to-length (TTL) coupling is another major limiting factor for the TianQin mission. The occurrence of TTL coupling is attributed to angular jitter and misalignments within the optical setup. It is documented that after integration, the expected TTL coupling coefficients for LISA, a space-based mission for the detection of gravitational waves spearheaded by the European Space Agency (ESA), reach 8.5 mm/rad~\cite{LISA2024}. Implementing the beam alignment mechanism (BAM) can partially reduce this coefficient to around 2.3 mm/rad~\cite{paczkowski2022postprocessing}. Unfortunately, despite implementing compensation, the expected TTL noise continues to exceed the defined noise threshold, thus requiring its removal during postprocessing. For TianQin, the TTL contribution is expected to stay below 3 mm/rad following beam realignment. The residual TTL noise, assigned post-processing, is anticipated to be 0.3 pm/Hz$^{1/2}$~\cite{wang2025postprocessing, chen2025tilt}.
\begin{figure}[ht] 
\centering
\includegraphics[width=0.48\textwidth]{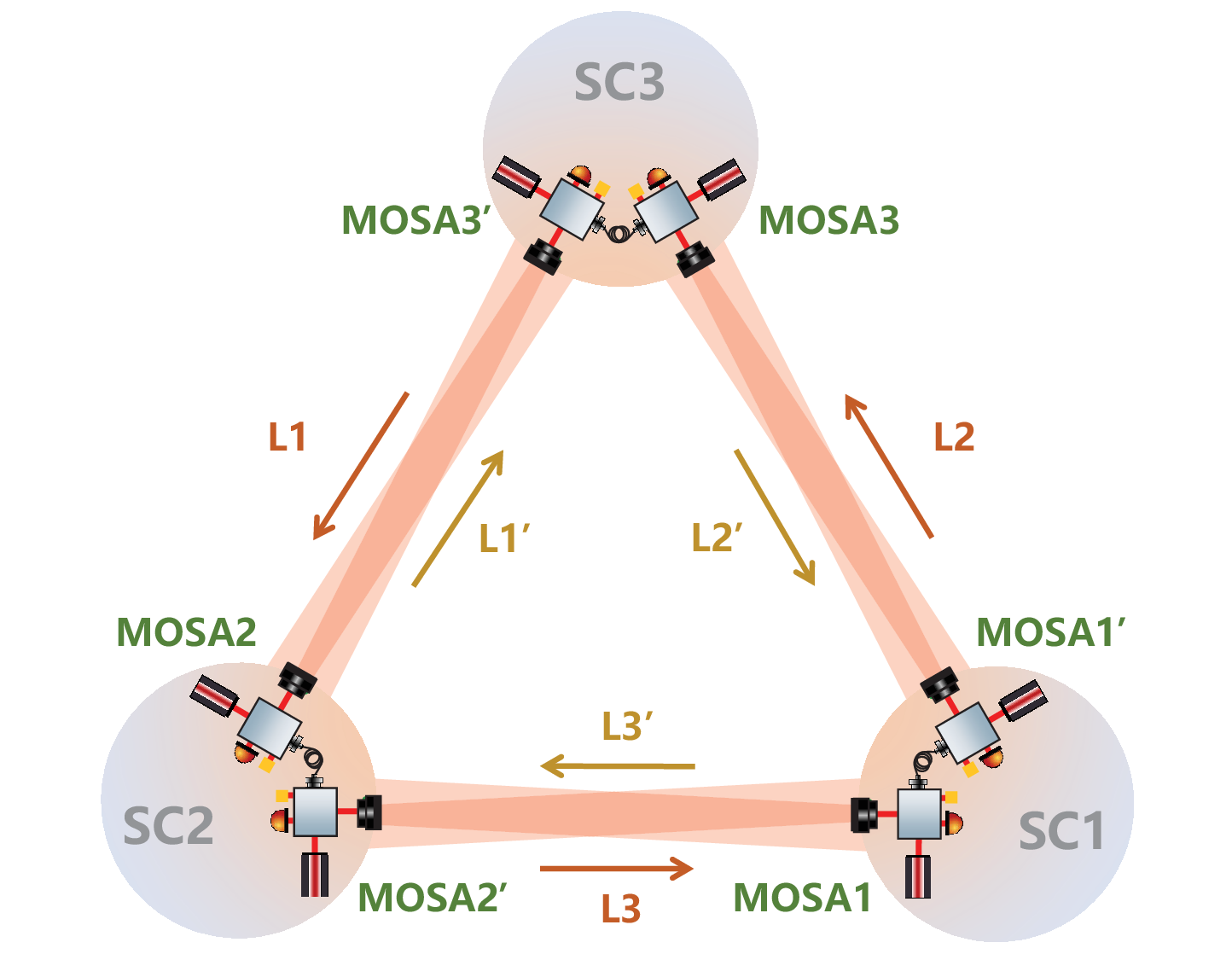}
\caption{\label{fig:constellationDemo} In the context of TDI labelling conventions, the MOSA located on the left side of satellite $i$ is referred to as MOSA$i$, while the corresponding MOSA on the right side is represented as MOSA$i'$. The arm length opposite to satellite $i$ is denoted as $\text{L}_i$. Arm lengths oriented clockwise are indicated using a prime, $\text{L}_{i'}$, whereas those oriented counter-clockwise retain the designation $\text{L}_{i}$. }  
\end{figure}

To enable a thorough examination of this type of noise, our study should begin with the basics of the satellite's payload design and its pointing strategy. Specifically, for TianQin, it will implement the two TMs and telescope pointing scheme, as described in Ref.\cite{fang2024payload}. Each TianQin satellite will contain a pair of movable optical subassemblies (MOSAs). Independent of one another, each MOSA encompasses an optical bench (OB), a telescope, and an inertial sensor that contains a free-falling test mass (TM). In the mission period, a vital component of payload management involves directing the outgoing laser beams towards remote satellites. The interferometric measurement requirement at the pm/Hz$^{1/2}$ scale necessitates a pointing accuracy of approximately 10 nrad for DC bias and 10 nrad/Hz$^{1/2}$ for jitters induced by the Drag-Free and Pointing Control System (DFPCS), which regulates essential maneuvers during the scientific operations. 

The principal concept of TTL noise reduction involves comprehending the angular misalignment. This misalignment, occurring between the wavefronts of interfering beams, is determined through the calculation of the differential phase, a method recognized as the differential wavefront sensing (DWS) technique~\cite{morrison1994automatic}. The intersatellite interferometer DWS readouts are considered as the total angular jitters induced by the satellite and MOSAs attitude control with respect to the incoming beam, which are used for TTL estimation in postprocessing without extra disturbance in the science operations. Previous studies have proposed an approach to estimate linear TTL coupling coefficients and their drifts using the null TDI channel $\zeta$ with the simplest combination $C_3^{12}$~\cite{wang2025postprocessing}. This scheme enables the estimation of the TTL coefficients for the subtraction of TTL noise in the Michelson variables $X$, $Y$, and $Z$ for further GWs analysis. However, the study did not account for a realistic jitter arising from the closed-loop dynamics, such as the dynamic interaction between the two MOSAs within the TTL estimation framework. Moreover, numerous studies (see Refs.\cite{george2023, houba2022optimal, houba2022lisa}) have demonstrated the critical importance of jitter correlation for TTL subtraction. 

However, examining the relationship between the three-axis jitters of the satellite and the yaw jitters in MOSAs represents an area not yet investigated by preceding TTL noise subtraction studies. Integrating more real dynamics data into the TTL noise analysis is crucial, particularly when employing least squares estimation, as it becomes markedly biased with elevated levels of readout noise. This bias is dependent on the jitter characteristics, and the DWS noise level \cite{hartig2025tilt}. Within its scientific data postprocessing framework, the LISA mission has already merged the drag-free closed-loop control data. This fusion seeks to create a more genuine noise generation process, thus allowing for the TDI effect's validation with the observed OB jitter, as elaborated in Refs.\cite{inchauspe2022new, heisenberg2023lisa}.

To further explore TTL noise subtraction using the null TDI channel, this research initially develops a comprehensive multi-body dynamics framework. This framework incorporates thruster configuration constraints, MOSA degrees of freedom, and DWS-based attitude determination, facilitating a thorough simulation of the DFPCS. This simulation examines the pointing control achieved through DWS-based attitude determination, alongside the synchronized control of DFPCS subsystems. We demonstrate TianQin DFPCS operating in science mode with closed-loop control, attaining the desired outcomes, while accurately providing the satellite and MOSAs' realistic jitters for the study of TTL subtraction. 

Based on closed-loop simulation data, we propose that employing a single-axis rotation control strategy and utilizing high-frequency band (0.1–1 Hz) data for estimating coupling coefficients would be advantageous for TTL noise suppression in TianQin. To improve estimation accuracy, we utilize various null TDI channels (see Refs.~\cite{muratore2020revisitation, muratore2022time, muratore2023effectiveness}) and suggest opting for sophisticated combinations like $C_3^{14}$ rather than simpler configurations like $C_3^{12}$. Additionally, introducing an artificially modulated MOSA jitter signal at 1 Hz would further improve parameter estimation precision while satisfying the control requirement.

The structure of this paper is laid out as follows, Sec.~\ref{sec: coordinate} provides a concise introduction to the coordinate systems and examines the jitter characteristics resulting from closed-loop control. In Sec.~\ref{sec: subtractionProcedure}, we present the TTL noise model alongside a subtraction strategy. Sections \ref{sec: setup} and \ref{sec: result} detail the numerical simulation setup and its corresponding results, respectively. Two methods aimed at enhancing the estimation accuracy are discussed in Sec.~\ref{sec: discussion}. Sec.~\ref{sec: conclusion} concludes the paper.


\section{Coordinate system and closed-loop dynamics}\label{sec: coordinate}
In this section, we introduce the foundational framework and dynamic modeling for the DFPCS, along with closed-loop simulations. We discuss the control residual noises, known as ``jitter'', which can exhibit some correlation between two MOSAs as a result of the control strategy.

\subsection{Coordinate system}
To simulate the system effectively, it's essential to track the satellite, TM, and MOSA dynamics. This requires setting up multiple non-inertial reference and body-fixed frames to represent the system's dynamics. As this study primarily concentrates on inter-satellite TTL noises, we only introduce the coordinates related to the satellites and MOSAs. The frames are outlined as follows.
\begin{figure}[ht]
\centering
\includegraphics[width=0.48\textwidth]{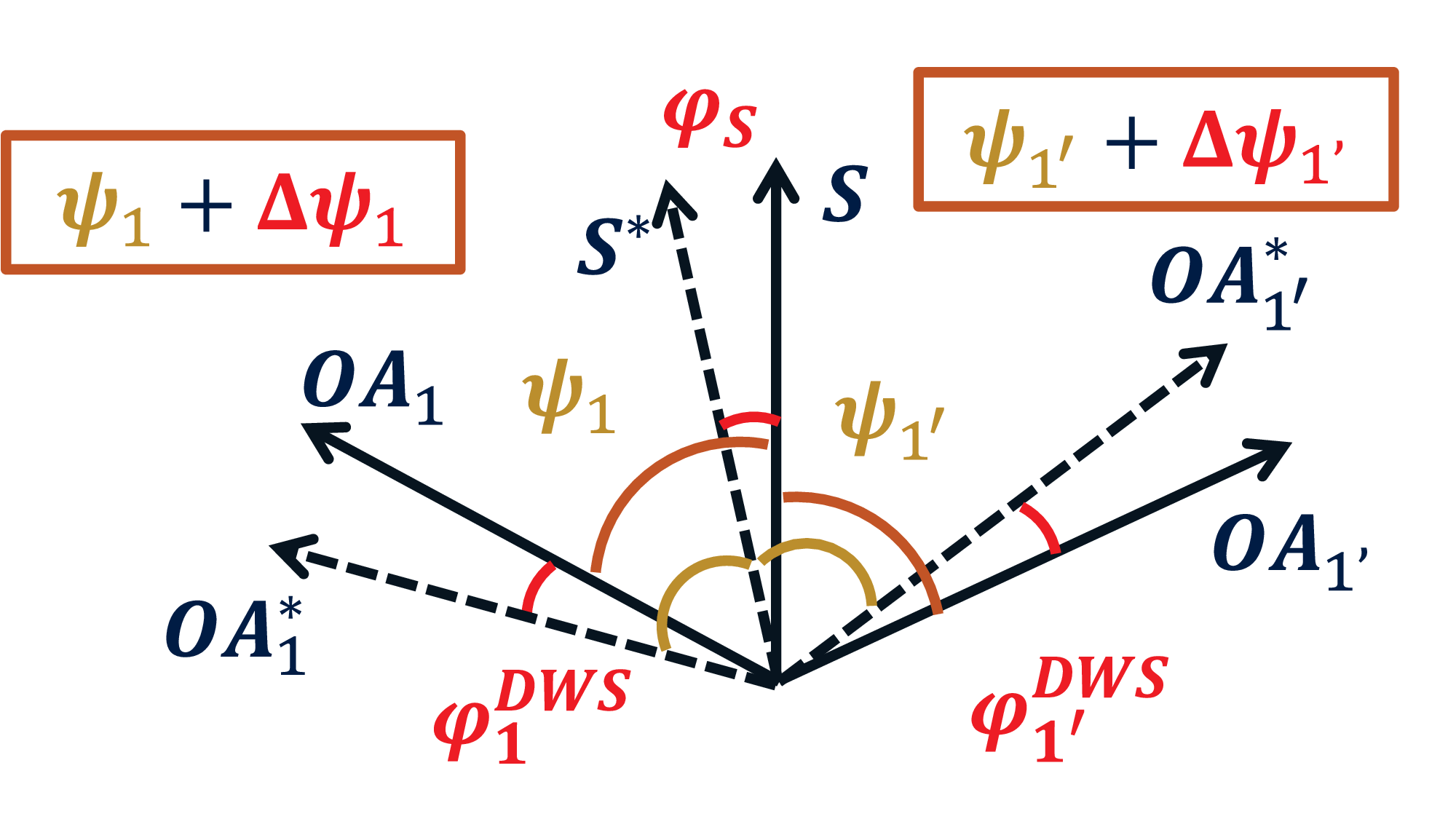}
\caption{\label{fig:jitterDemo} The schematic diagram briefly illustrates the angular jitter between the coordinates within the constellation plane, where the angles $\psi_{OA_1}$ and $\psi_{OA_{1'}}$ is the MOSAs opening angle for compensation of the breathing angle. The true jitter in the DWS readout is represented by $\varphi_{DWS_1}$ and $\varphi_{DWS_{1'}}$. The satellite jitters, along with the two MOSAs, are denoted by $\varphi_S$, $\Delta\psi_1$, and $\Delta\psi_{1'}$. }
\end{figure}

\subsubsection{Body-fixed frames}
\begin{itemize}
   \item The $\mathcal{I}$ frame is the inertial frame utilized to detail satellite motion in geocentric orbits. This paper employs the J2000-based Earth-centered equatorial coordinate system with axes $\vec{X}_\mathrm{I}$, $\vec{Y}_\mathrm{I}$, and $\vec{Z}_\mathrm{I}$.
   
   \item The $\mathcal{S}_i$ frame, where $i=1,2,3$, is firmly connected to a satellite and represents its motion. This is constructed as follows.
   \begin{itemize}
       \item The origin is at the CoM of the satellite. 
       \item $\vec{X}_{\mathrm{S}_i}$ bisects the $60^{\circ}$ angle between the two optical assemblies. 
       \item $\vec{Z}_{\mathrm{S}_i}$ is perpendicular to the solar panel. 
       \item $\vec{Y}_{\mathrm{S}_i}$ is determined by the right-hand rule. 
   \end{itemize}
   
   \item The $\mathcal{OA}_i$ frame characterizes the movement of a single MOSA, which is also referred to as the optical assembly.
   \begin{itemize}
       \item The origin is at the geometric center of the electrode housing. 
       \item $\vec{X}_{\mathrm{OA}_i}$ aligns with the optical axis of the MOSA. 
       \item $\vec{Z}_{\mathrm{OA}_i}$ aligns with $\vec{Z}_\mathrm{S}$
       \item $\vec{Y}_{\mathrm{OA}_i}$ is determined by the right-hand rule. 
   \end{itemize}
\end{itemize}

\subsubsection{Reference frames}
The reference frame is crucial for setting the DFPCS control goals. Based on the 20 DoFs plant, pointing control aims to align MOSAs' optical axes with the two incoming laser beams. Hence, the target frame for pointing control is defined as follows
\begin{itemize}
    \item The $\mathcal{OA}_i^*$ frame describes the target/nominal attitude of one optical assembly. The attitude of $\mathcal{OA}_i^*$ frame relative to the $\mathcal{O}_i$ frame is defined as $\left( {{\theta_i^\text{DWS}},{\eta_i^\text{DWS}},{\varphi_i^\text{DWS}}} \right)$, using the XYZ sequence.
    \begin{itemize}
        \item $\vec{X}_{\mathrm{OA}_i^*}$ is the unit vector opposite to the direction of the corresponding incoming laser beams.
        \item $\vec{Z}_{\mathrm{OA}_i^*}$ is normal to the plane formed by $\vec{X}_{\mathrm{OA}_i^*}$ and $\vec{X}_{\mathrm{OA}_{i'}^*}$.
        \item $\vec{Y}_{\mathrm{OA}_i^*}$ is deduced from the cross product of $\vec{X}_{\mathrm{OA}_i^*}$ and $\vec{Z}_{\mathrm{OA}_i^*}$.
        \item The origin coincides with the origin of $\mathcal{OA}_i$ frame.
    \end{itemize}
\end{itemize}
The ${\eta_l^\text{DWS}}$ and ${\varphi_l^\text{DWS}}$ respectively represent the pitch and yaw angles of the $\mathcal{OA}_l^*$ relative to the $\mathcal{OA}_l$ frame. These serve as the primary attitude data during the science mode. The objective of pointing control is to reduce these values below the specified threshold. Nevertheless, since MOSA is only capable of supporting yaw motion, it is inadequate to meet the pointing control objective on its own. Thus, effective pointing control necessitates the synchronized execution of both satellite attitude control and MOSA pointing management. To control the attitude of a satellite, it is essential to establish the appropriate target reference frame for the satellite. The objective is to align the satellite's $\mathcal{S}$ frame with the designated target reference frame $\mathcal{S}^*$. It is important to note that the satellite target reference frame definition varies, contingent on MOSA's motion mode. Below is a commonly used definition for symmetric MOSA rotation.
\begin{itemize}
    \item The $\mathcal{S}_i^*$ frame ($i=1,2,3$) describes the target/nominal attitude for one satellite. The attitude of $\mathcal{S}_i$ frame relative to the $\mathcal{S}_i^*$ frame is defined as (${\theta_{S_i}},{\eta_{S_i}},{\varphi_{S_i}}$), using the XYZ sequence.
    \begin{itemize}
        \item $\vec X_{S_i^*}$ is the bisector of the angle between $x_{OA_1}^*$  and $x_{OA_2}^*$.
        \item $\vec Z_{S_i^*}$ is parallel to $z_{OA_i}^*$.
        \item $\vec Y_{S_i^*}$ is deduced from the cross product of $x_{S}^*$ and $z_{S}^*$.
        \item The origin coincides with the origin of $\mathcal{S}$-Frame.
    \end{itemize}

   \item The $\mathcal{S}_i^*$ frame ($i=1,2,3$) describes the target/nominal attitude for one satellite. 
   \begin{itemize}
       \item The origin coincides with the one of $\mathcal{S}_i$. 
       \item $\vec X_{S_i^*}$ point towards the incenter of the triangular constellation. 
       \item $\vec Z_{S_i^*}$ is orthogonal to the constellation plane.
       \item $\vec Y_{S_i^*}$ is built from the cross product of the two above.
   \end{itemize}
\end{itemize}

\subsection{Closed-loop dynamics}
Satellite pointing jitter is closely associated with the DFPC approach. The orientation of the satellite is largely affected by thruster disturbances, while MOSA operates through onboard piezoelectric ceramic devices. Therefore, to examine the jitter in both the satellite and MOSA, it's essential to conduct a comprehensive study from simulations of closed-loop control dynamics.

\subsubsection{Control scheme}
This section focuses primarily on the control scheme of the DFPCS. As for the satellite dynamics model (one can see in Ref.\cite{zhang2024nonlinear}), while important, it is not the central emphasis of this work, and is therefore not discussed in detail here. In science mode, the DFPCS is required to accomplish the following essential tasks.
\begin{itemize}
    \item Drag-Free Control, which Precisely maneuvers the satellite using thrusters to follow two test masses along two sensitive directions, ensuring that the test masses experience minimal drag along these axes. The current design requirement specifies that the amplitude spectral density (ASD) of the relative displacement of test masses along the sensitive axis must be below 1.6 nm/Hz$^{1/2}$.

    \item Pointing control, which adjusts the satellite's attitude to align it with the constellation's rotation while modulating the MOSA to account for changes in the breathing angle, securing the laser link. This necessitates synchronized control over both satellite orientation and MOSA aiming without affecting drag-free operation. Achieving alignment between local telescopes and incoming laser beams from remote satellites demands pointing stability of 10 nrad for the DC bias and 10 nrad/Hz$^{1/2}$ for jitters.

    \item Electrostatic Suspension control, which applies electrostatic control to the remaining DoFs of the TMs via the inner electrodes of the housing. These forces are necessary to counteract any differential acceleration between the two TMs that the drag-free control cannot compensated. It is essential to restrict the suspension force to reduce actuation noise and unwanted force gradients in the housing, especially in the event of cross-coupling with the sensitive axes.
\end{itemize}

Using the attitude determination equation presented in Sec.~\ref{subsec:jitterAnalysis}, we establish a stable link between sensor data and control outputs for targeting control. The approach for decoupling non-orthogonal sensitive axes in drag-free control is thoroughly covered in~\cite{xiao2022drag}, and will not be repeated here.

\subsubsection{Jitter characteristics}\label{subsec:jitterAnalysis}
This section explains the process of using the DWS readouts to determine satellite and MOSAs attitudes. The connection between attitudes and DWS readouts can be established through the coordinate transformation matrix. For example, transformation from the $\mathcal{OA}_1^*$ frame to the $\mathcal{S}$ frame can be achieved in two ways: a) transforming from the $\mathcal{OA}_1^*$ frame to the $\mathcal{OA}_1$ frame, followed by transformation to the $\mathcal{S}$ frame (the left-hand side of Eq.~\eqref{eq:transform}), and b) transforming from the $\mathcal{OA}_1^*$ frame to the $\mathcal{S}^*$ frame, then to the $\mathcal{S}$ frame (the right-hand side of Eq.~\eqref{eq:transform}). Now we can establish the relative jitters relationship between the four coordinates: $\mathcal{OA}_1$, $\mathcal{OA}_1^*$, $\mathcal{S}$, and $\mathcal{S}^*$. It can be summarized by the following equation,
\begin{equation}\label{eq:transform}
R_{OA_1}^S ~ R_{OA_1^*}^{OA_1}  = R_{S^*}^S ~ R_{OA_1^*}^{S^*}.
\end{equation}

The comprehensive relationships of the coordinate transformation matrix, alongside their derivation, are illustrated in Fig.~\ref{fig:jitterDemo} and discussed in Appendix.\ref{sec: AppendixA}. In conclusion, the connection between attitude and DWS output is elucidated in Eq.~\eqref{eq:jitter2DWSFull}, which can alternatively be expressed as
\begin{widetext}
\begin{equation}\label{eq:jitter2DWS}
    \begin{aligned}
    \begin{bmatrix}
    \eta_1^\text{DWS} \\ 
    \varphi_1^\text{DWS} \\ 
    \eta_2^\text{DWS} \\ 
    \varphi_2^\text{DWS}
    \end{bmatrix} 
    & = \begin{bmatrix}
    - \sin(\psi_{1}) & \cos(\psi_{1}) & 0 & 0 & 0 \\
    0 & 0 & 1 & -1 & 0 \\
    - \sin(\psi_{2}) & \cos(\psi_{2}) & 0 & 0 & 0 \\
    0 & 0 & 1 & 0 & -1
    \end{bmatrix}
    \begin{bmatrix}
    \theta_\text{S} \\ 
    \eta_\text{S} \\ 
    \varphi_\text{S} \\ 
    \Delta\psi_1 \\ 
    \Delta\psi_2
    \end{bmatrix} \approx \begin{bmatrix}
    -\frac{1}{2} & \frac{\sqrt{3}}{2} & 0 & 0 & 0 \\
    0 & 0 & 1 & -1 & 0 \\
    \frac{1}{2} & \frac{\sqrt{3}}{2} & 0 & 0 & 0 \\
    0 & 0 & 1 & 0 & -1
    \end{bmatrix}
    \begin{bmatrix}
    \theta_\text{S} \\ 
    \eta_\text{S} \\ 
    \varphi_\text{S} \\ 
    \Delta\psi_1 \\ 
    \Delta\psi_2
    \end{bmatrix}
    \end{aligned}.
\end{equation}
\end{widetext}
The values of $|\psi_1|$ and $|\psi_2|$ are each 30$^\circ$ $\pm$ 0.1$^\circ$, attributed to constellation breathing, and they are opposite in nature. This is a result of the MOSA coordinates rotating in different directions—one moves counterclockwise while the other moves clockwise—relative to the satellite's coordinate system. DWS readouts free of measurement noise are represented by $\eta_1^\text{DWS}$, while those encompassing sensing noise are denoted as $\hat{\eta}_1^\text{DWS}$. The resulting measurements acquired via the DWS technique can be expressed as
\begin{subequations}\label{eq:DWS_meas}
\begin{align}
    \hat{\eta}_i^\text{DWS} &= {\eta}_i^\text{DWS} + \delta{\eta}_i, \\
    \hat{\varphi}_i^\text{DWS} &= {\varphi}_i^\text{DWS} + \delta{\varphi}_i,
\end{align}
\end{subequations}
where $\delta{\eta}_i$ and $\delta{\varphi}_i$ are the sensing noise introduced in the ISI after performing calibration, accounting for the magnification $M$ of the telescope and OB.

In a practical detection scenario, only knowing the DWS measurement details for 4 angles makes it impossible to distinctly identify the information for 5 angles. Therefore, setting constraints is a prerequisite for using Eq.~\eqref{eq:OA2S}. This article considers the two most distinctive modes of MOSA motion, symmetric MOSA actuation and single MOSA actuation.
\begin{itemize}
\item Single MOSA actuation.
\end{itemize}
In the scenario of a single MOSA actuation, we introduced the subsequent constraint relationships as
\begin{equation}\label{eq:single}
    \Delta{\psi_2} = 0.
\end{equation}
Substituting \eqref{eq:single} into \eqref{eq:jitter2DWS}, the attitude determination matrix for transforming DWS readout under the single MOSA actuation mode is given by
\begin{equation}\label{eq:singleTransform}
    \left[ {\begin{array}{*{20}{c}}
    {{\theta_\text{S}}}\\
    {{\eta_\text{S}}}\\
    {{\varphi_\text{S}}}\\
    {\Delta{\psi_1}}
    \end{array}} \right] = \left[ {\begin{array}{*{20}{c}}
    -1 & 0 & 1 & 0\\
    \frac{1}{\sqrt{3}} & 0 &\frac{1}{\sqrt{3}}&0\\
    0 & 0 & 0 & 1\\
    0 &-1 & 0 & 1
    \end{array}} \right] \left[{\begin{array}{*{20}{c}}
    {\hat{\eta}_1^\text{DWS}}\\
    {\hat{\varphi}_1^\text{DWS}}\\
    {\hat{\eta}_2^\text{DWS}}\\
    {\hat{\varphi}_2^\text{DWS}}
    \end{array}}\right].
\end{equation}

\begin{itemize}
    \item Symmetric MOSA actuation.
\end{itemize}
In the scenario of symmetric MOSA actuation, we have introduced the following constraint relationship as
\begin{equation}\label{eq:symmetric}
\Delta{\psi} = \Delta{\psi_1} = - \Delta{\psi_2}.
\end{equation}
By inserting \eqref{eq:symmetric} into \eqref{eq:jitter2DWS}, the resulting matrix for attitude determination under symmetric MOSA actuation mode is
\begin{equation}\label{eq:symTransform}
\left[ {\begin{array}{*{20}{c}}
{{\theta_\text{S}}}\\
{{\eta_\text{S}}}\\
{{\varphi_\text{S}}}\\
\Delta{\psi}
\end{array}} \right] = \left[ {\begin{array}{*{20}{c}}
-1 & 0 & 1 & 0\\
{\frac{1}{\sqrt3}} & 0 & {\frac{1}{\sqrt3}} & 0\\
0 & {\frac{1}{2}}& 0 & {\frac{1}{2}}\\
0 & {-\frac{1}{2}}& 0 & {\frac{1}{2}}
\end{array}} \right]\left[ {\begin{array}{*{20}{c}}
{\hat{\eta}_1^\text{DWS}}\\
{\hat{\varphi}_1^\text{DWS}}\\
{\hat{\eta}_2^\text{DWS}}\\
{\hat{\varphi}_2^\text{DWS}}
\end{array}} \right].
\end{equation}

\begin{figure}[ht]
\begin{minipage}{0.48\textwidth}
  \centering
  \includegraphics[width=\textwidth]{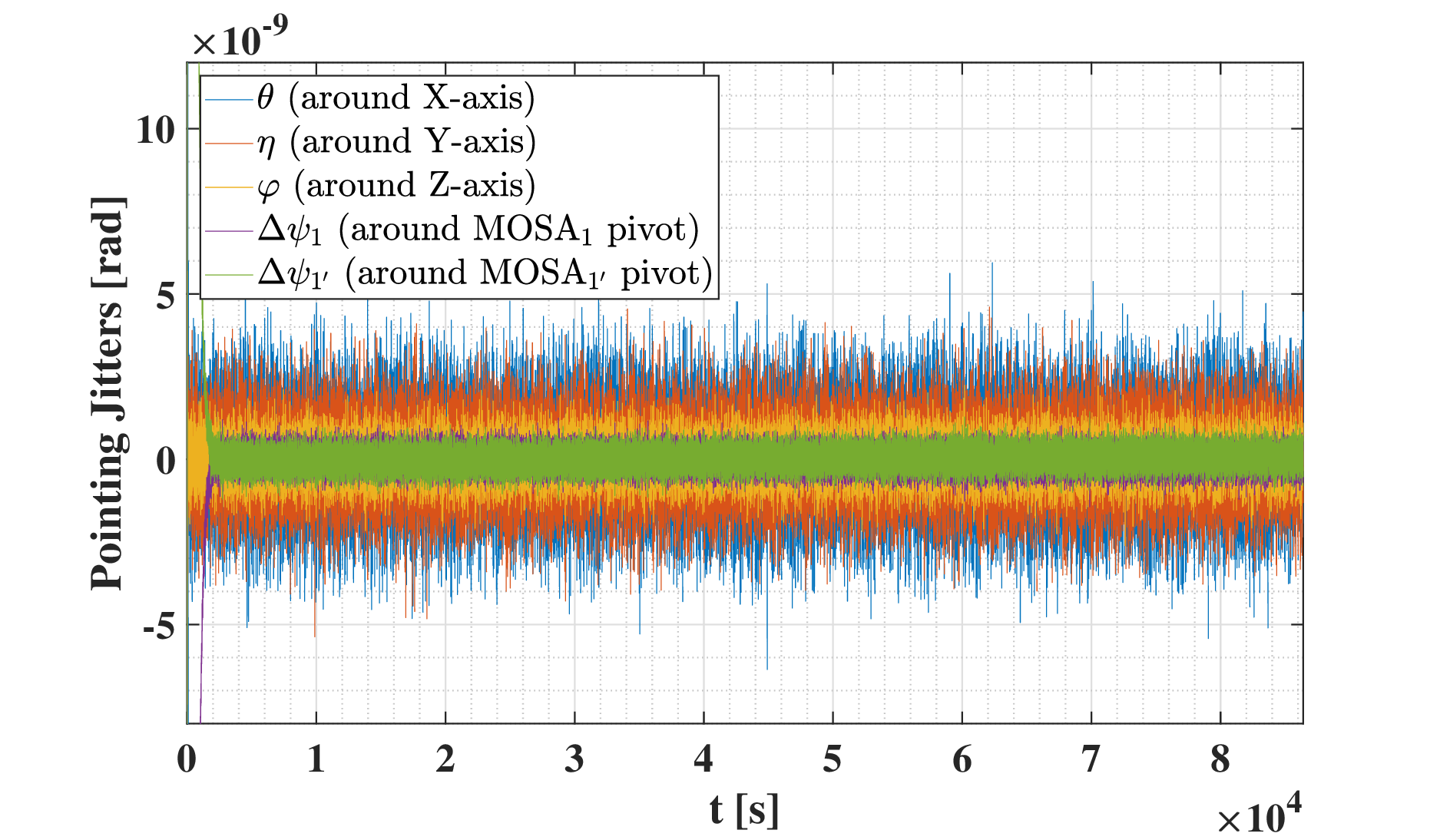}
\end{minipage}
\begin{minipage}{0.48\textwidth}
  \centering
  \includegraphics[width=\textwidth]{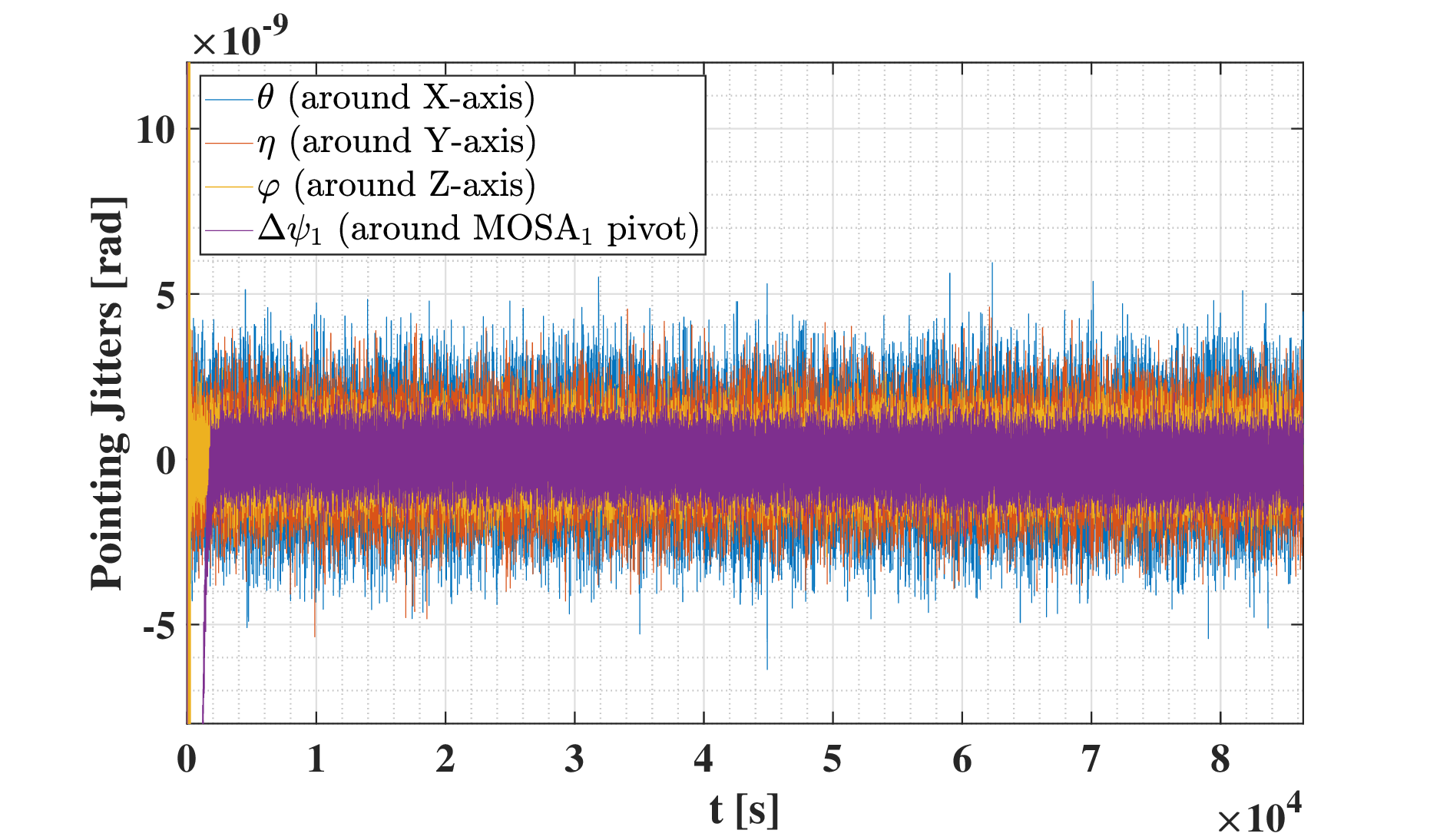}
\end{minipage}
\caption{\label{fig:Jitter_timedomain} Time-domain jitter simulation diagrams under two different control strategies. The upper plot illustrates the symmetrical rotation of two MOSAs to compensate for breathing angles, whereas the lower plot demonstrates the rotation of a single MOSA, which exhibits relatively larger dynamic ranges.}
\end{figure}

To conclude, Eq.~\eqref{eq:jitter2DWS} outlines a simple linear correlation between DWS readouts and the satellite pointings. The derived equations, Eq.~\eqref{eq:singleTransform} and \eqref{eq:symTransform}, are crucial as they enable the determination of both satellite attitude and MOSA pointing. Through closed-loop dynamics, we can simulate the time-domain data of real jitter as shown in Fig.~\ref{fig:Jitter_timedomain}. From the two simple linear formulas, it can be observed that the jitter compensation required for the two MOSAs on the same satellite depends on their rotation modes. In cases of symmetrical rotation, the jitter adjustments for both MOSAs are equal in size but opposite in direction, which might lead to a strong correlation in their actual jitters. However, for single-axis rotation, as the other MOSA remains stationary, there is no significant coupling relationship between them. Yet, the compensation required for the single MOSA is considerably larger. This finding will also be confirmed in Sec.~\ref{sec: jitterResult} in the frequency domain.


\section{TTL noise subtraction procedure} \label{sec: subtractionProcedure}
\begin{figure*}[ht]
\centering
\includegraphics[width=1\textwidth]{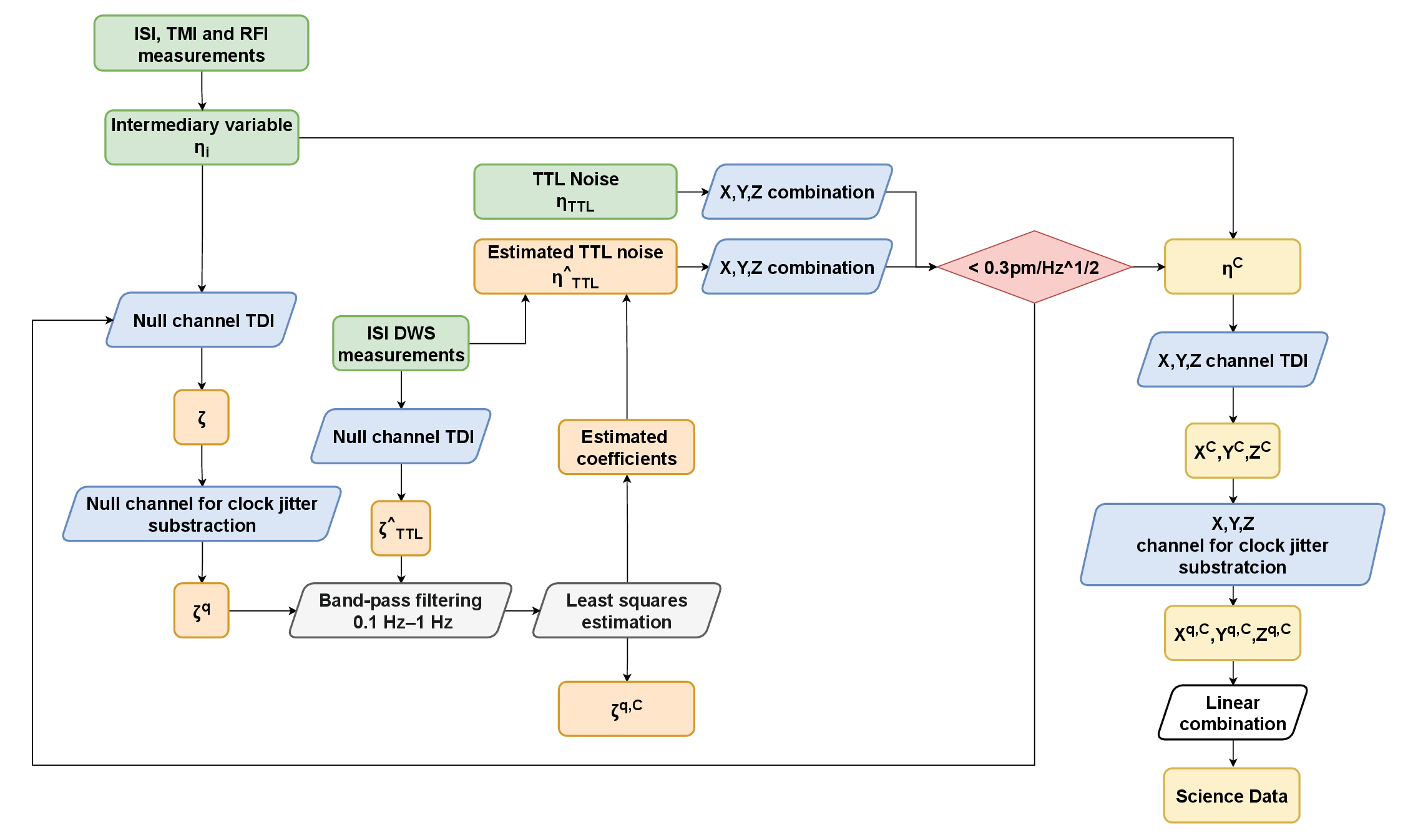}
\caption{\label{fig:Demo_procedure} Illustrative diagram of the null TDI channel method for TTL noise subtraction. By employing the null channel method, we can directly estimate all 48 TTL coefficients, including the linear drifts. To verify the accuracy of the calibration coefficients, it is necessary to examine whether the residual noise remains below 0.3 pm/Hz$^{1/2}$ using the X, Y, Z combination. }
\end{figure*}

In this section, we propose the utilization of the null channel to mitigate laser phase noise and GW signals prior to estimating the TTL coefficients. These coefficients will be utilized to subtract the TTL noise from the intermediate variable $\eta_i$. Subsequently, we will evaluate the remaining noise in $X$ to determine if it meets the criterion of 0.3 pm/Hz$^{1/2}$. Finally, we are able to construct other TDI variables for GW extraction, e.g., $X$, $Y$, and $Z$, that are free from TTL noise as illustrated in Fig.~\ref{fig:Demo_procedure}.

\subsection{TTL noise modeling in $\eta$-terms}
Here, we shall conduct a brief analysis on the procedure for defining the intermediate variable $\eta_i$ as well as on the modeling of TTL noise~\cite{Wanner2024, wang2025postprocessing}. The TianQin constellation and the labeling notations for TDI used in this paper are shown in Fig.~\ref{fig:constellationDemo}. For the TDI notation employed in this paper, please refer to the accompanying Fig.~\ref{fig:constellationDemo}. The carrier-to-carrier interferometric measurements of ISI, TMI, and RFI can be written as \cite{otto2012tdi, tinto2018time}
\begin{subequations}\label{eq:carrier_meas}
\begin{align}
\text{isi}_i & = p_i - {\bf{D}}_{k}p_{j} + h_{k} 
+ \frac{\nu_{j}}{c}\left(-\boldsymbol{e}_{k'}\cdot\boldsymbol{\Delta}_i
-\boldsymbol{e}_{k}\cdot{\bf{D}}_{k}\boldsymbol{\Delta}_{j}\right) \notag\\
&\quad + n_{k,\text{isi}}^{\text{TTL}} + n_{i,\text{isi}}^{\text{opt}} 
+ n_{i,\text{isi}}^{\text{ro}} - {a_i}{q_i},\\[0.5em]
\text{tmi}_i &= p_i - p_{i'} - \frac{2\nu_{i}}{c}
\left(\boldsymbol{e}_{k'}\cdot\boldsymbol{\Delta}_i
-\boldsymbol{e}_{k'}\cdot\boldsymbol{\delta}_i\right) \notag\\
&\quad - n_{i,\text{tmi}}^{\text{TTL}} + n_{i'i}^{\text{bl}} 
+ n_{i,\text{tmi}}^{\text{opt}} + n_{i,\text{tmi}}^{\text{ro}}  
- {b_i}{q_i},\\[0.5em]
\text{rfi}_i &= p_i - p_{i'} + n_{i'i}^{\text{bl}} 
+ n_{i,\text{rfi}}^{\text{opt}} + n_{i,\text{rfi}}^{\text{ro}} - {b_i}{q_i}.
\end{align}
\end{subequations}
Notations $c$, $\nu$, $p$, $q$, $h$, $\boldsymbol{\Delta}$, $\boldsymbol{\delta}$, $n^{\text{TTL}}$, $n^{\text{bl}}$, $n^{\text{opt}}$, $n^{\text{ro}}$ denote the speed of light in vacuum, laser frequency, laser phase noise, clock phase noise, GW signal, satellite translational jitter, TM translational jitter, TTL noise, backlink fiber noise, optical polarization and stray light noise, readout noise, respectively. $\boldsymbol{e}_{k}$ denotes the unit vector pointing from MOSA$_j$ to MOSA$_i$. For any signal $x(t)$, we define the delay operator ${\bf{D}}_{k}$ as
\begin{equation}
	{\bf{D}}_{k}x(t) = x(t - d_{k}(t))
\end{equation}
where $d_{k}$ denotes the light travel time between MOSA$j'$ and MOSA$i$, with $i,j,k\in\{1,1',2,2',3,3'\}$, $i \neq j \neq k$. We also introduce the advancement operator ${\bf{A}}_{k}$, the inverse of ${\bf{D}}_{k}$, given by
\begin{equation}
	{\bf{A}}_{k}x(t) = x(t + {d}_{k}(t))
\end{equation}
Both operators will be used in the following text.

The principal objective of TDI is to mitigate the excessive laser phase noise through the creation of a virtual interferometer. Initially, it is essential to eliminate the OB jitter along the sensitive axis by employing the TMI and RFI measurements, i.e.,
\begin{equation}\label{eq:xi}
	\xi_i = \text{isi}_i - \frac{1}{2}\left(\frac{\nu_j}{\nu_i}(\text{tmi}_i-\text{rfi}_i)+{\bf{D}}_{k}(\text{tmi}_{j'}-\text{rfi}_{j'})\right).
\end{equation}
The subsequent step involves substituting the primed laser phase noise with the unprimed one, such as
\begin{subequations}\label{eq:eta}
\begin{align}
    \eta_i &= \xi_i - \frac{1}{2}{\bf{D}}_{k}(\text{rfi}_j-\text{rfi}_{j'}), \\
    \eta_{i'} &= \xi_{i'} - \frac{1}{2}(\text{rfi}_{i'}-\text{rfi}_i).
\end{align}
\end{subequations}
By cyclic permutation of the indices $i\rightarrow{j}\rightarrow{k}\rightarrow{i}$, one can compute the remaining four intermediary variables $\eta_{j,j',k,k'}$. 

In the context of modeling TTL noise, it is commonly represented by its length. Therefore, we articulate the TTL noise in the three interferometers as follows
\begin{subequations}\label{eq:eta^TTL}
\begin{align}
    \eta_{k,\text{isi}}^{\text{TTL}} &= \frac{\nu_{j}}{c}(\rho_{i,\text{isi}}^{\text{RX}}+{\bf {D}}_{k}\rho_{j',\text{isi}}^{\text{TX}}), \\
    \eta_{i,\text{tmi}}^{\text{TTL}} &= \frac{\nu_{i}}{c}\rho_{i,\text{tmi}}, \\
    \eta_{i,\text{rfi}}^{\text{TTL}} &= 0, 
\end{align}
\end{subequations}
where $\rho_{i,\text{isi}}^{\text{RX}}$ and $\rho_{{j'},\text{isi}}^{\text{TX}}$ denote the TTL noise in the ISI received (RX) and transmitted (TX) beam, respectively, and $\rho_{i,\text{tmi}}$ denotes the TTL noise in TMI. By substituting Eq.~(\ref{eq:eta^TTL}) into Eq.~(\ref{eq:carrier_meas}) and subsequently Eq.~(\ref{eq:xi}), we derive the ultimate TTL contributions within the single-link readout, expressed as
\begin{align}\label{eq:xi^TTL}
	\xi_i^{\text{TTL}} = \frac{\nu_{j'}}{c}\left(\rho_i^{\text{RX}}+{\bf{D}}_{k}\rho_{j'}^{\text{TX}}\right),
\end{align}
with
\begin{subequations}
	\label{rho_def}
	\begin{align}
		\rho_i^{\text{RX}} &= \rho_{i,\text{isi}}^{\text{RX}}+\frac{1}{2}\rho_{i,\text{tmi}}, \\
		\rho_{j'}^{\text{TX}} &= \rho_{j',\text{isi}}^{\text{TX}}+\frac{1}{2}\rho_{j',\text{tmi}}.
	\end{align}
\end{subequations}
According to Eqs. \eqref{eq:eta}, \eqref{eq:xi^TTL} and \eqref{eq:eta^TTL}, we directly have $\eta_i^{\text{TTL}}=\xi_i^{\text{TTL}}$. This noise, referred to as tilt-to-length noise, is calculated using the coupling coefficients and the angular jitters $({\theta}_i^\text{DWS}, {\eta}_i^\text{DWS}, {\varphi}_i^\text{DWS})$, commonly termed as the roll, pitch, and yaw angles of the MOSAs, and their relationship is detailed in the Appendix \ref{sec: AppendixA} as Eq.~\eqref{eq:jitter2DWSFull} shown.

Regarding TTL noise, the roll angle ${\theta}_i^\text{DWS}$ holds less significance, as minor rotations about the axis of the incoming beam result in negligible TTL noise. Therefore, the TTL contributions $\rho_i^{\text{RX}}$ and $\rho_{j'}^{\text{TX}}$ are expressed with the pitch and yaw angles, i.e.,
\begin{subequations}\label{eq:eta^TTLC}
\begin{align}
    \rho_i^{\text{RX}} &= C_{i,\eta}^{\text{RX}}{\eta}_i^\text{DWS} + C_{i,\varphi}^{\text{RX}}{\varphi}_i^\text{DWS}, \\
    \rho_{j'}^{\text{TX}} &= C_{j',\eta}^{\text{TX}}{\eta}_{j'}^\text{DWS} + C_{j',\varphi}^{\text{TX}}{\varphi}_{j'}^\text{DWS}.
\end{align}
\end{subequations}
Based on the discussion, the TTL noise $\eta_{k}^{\text{TTL}}$ in $\eta_i$ reads
\begin{align}\label{eq:eta_k^TTL}
    \eta_{k}^{\text{TTL}} &= \frac{\nu_j}{c}\left[C_{i,\eta}^{\text{RX}}{\eta}_i^\text{DWS} + C_{i,\varphi}^{\text{RX}}{\varphi}_i^\text{DWS}\right. \nonumber\\
    &\quad \left.+ {\bf{D}}_{k}\left(C_{j',\eta}^{\text{TX}}{\eta}_{j'}^\text{DWS} + C_{j',\varphi}^{\text{TX}}{\varphi}_{j'}^\text{DWS}\right)\right].
\end{align}
With the same procedure we get all the TTL noise $\eta^\text{TTL}$-terms in the $\eta$-terms, i.e., $\eta_{k}^{\text{TTL}}$ in $\eta_i$, $\eta_{j'}^{\text{TTL}}$ in $\eta_{i'}$, $\eta_{i}^{\text{TTL}}$ in $\eta_j$, $\eta_{k'}^{\text{TTL}}$ in $\eta_{j'}$, $\eta_{j}^{\text{TTL}}$ in $\eta_k$, $\eta_{i'}^{\text{TTL}}$ in $\eta_{k'}$. Propagating Eq.~\eqref{eq:eta_k^TTL} through a specific TDI combination yields the TTL contributions in the corresponding TDI variable. For instance, substituting $\eta_i$ and $\eta_{k}^{\text{TTL}}$ within any TDI combination, such as $X$, allows us to obtain the TDI variable $X$ and the TTL noise 
\begin{equation}\label{eq:realX^TTL}
{X}^{\text{TTL}} = \sum_{i,\alpha,N}{C}_{i,\alpha}^{N}{X}_{i,\alpha}^{N},
\end{equation}
with the symbols $\alpha = \eta,\varphi$ and $N = \text{RX},\text{TX}$.

\subsection{TTL residual noise modeling}
In order to mitigate TTL noise the actual angular jitters remain unknown, with only the measured jitters, as acquired via the DWS method, being accessible, as demonstrated in Eq.~\eqref{eq:DWS_meas}. In this study, we utilize the TDI configuration $\zeta$ to demonstrate the process of estimating and subtracting TTL noise, ultimately formulating a model for the residual noise. It is critical to estimate the linear coupling coefficients of the TTL, which are likely to demonstrate slow changes during flight. Although linear drift is not our primary concern, for completeness, we do consider the possible linear drift of the TTL coefficients during estimation. 

To estimate the TTL coefficients, according to Eq.~(\ref{eq:DWS_meas}), we need to propagate each measured angular jitter with sensing noise through $\zeta$ (e.g. Eq.~\eqref{eq:zeta_C312}) combination. The output is written as $\hat{\zeta}_{i,\alpha}^{N}$, which can be divided into the following two parts
\begin{equation}
\hat{\zeta}_{i,\alpha}^{N} = \zeta_{i,\alpha}^{N} + \delta \zeta_{i,\alpha}^{N}~,
\end{equation}
where the $\zeta_{i,\alpha}^{N}$ is the true jitter's TDI variable we set up in the simulation and the $\delta \zeta_{i,\alpha}^{N}$ is the DWS readout noise after the null channel. The estimated TTL noise in the $\zeta$ can now be given by 
\begin{align}\label{eq:zetahat^TTL}
\hat{\zeta}^{\text{TTL}} = \sum_{i,\alpha,l}\hat{C}_{i,\alpha}^{N}\hat{\zeta}_{i,\alpha}^{N}.
\end{align}
The TTL estimation is executed after clock noise suppression, so the real TDI variable can be written as $\zeta_\text{corr}^q$, we will perform least squares estimation for the TTL coupling coefficients $\hat{C}_{i,\alpha}^{N}$ with the $\zeta_\text{corr}^q$ and $\hat{\zeta}^{\text{TTL}}$. As a result, the estimated coefficient can be written as
\begin{equation}
\hat{C}_{i,\alpha}^{N} = C_{i,\alpha}^{N} + \delta{C}_{i,\alpha}^{N},
\end{equation}
where the $C_{i,\alpha}^{N}$ is the true coefficients we set up in the simulation and we use $\delta{C}_{i,\alpha}^{N}$ to denote the estimation error of the TTL coefficient. Then the estimated TTL noise $\hat{X}^{\text{TTL}}$ reads
\begin{align}\label{eq:Xhat^TTL}
\hat{X}^{\text{TTL}} = \sum_{i,\alpha,N}(C_{i,\alpha}^{N} + \delta C_{i,\alpha}^{N})(X_{i,\alpha}^{N} + \delta X_{i,\alpha}^{N}).
\end{align}
where the $X_{i,\alpha}^{N} + \delta X_{i,\alpha}^{N}$ is the measured angular jitter with sensing noise through X (e.g. Eq.~\eqref{eq:X}) combination. 

Finally, the $X_\text{corr}^{\text{TTL}}$ is the corrected TDI variable for $X$ as the TTL noise has been removed by the estimated one $\hat{X}^{\text{TTL}}$. Using Eq.~\eqref{eq:realX^TTL} and Eq.~\eqref{eq:Xhat^TTL} the TTL estimation error in $X_\text{corr}^{\text{TTL}}$ can be computed as
\begin{align}\label{eq:deltaX_corr^TTL}
    \delta X^{\text{TTL}} = - \sum_{i,j,l}\left(\delta{C}_i^{j,l}X_i^{j,l} + {C}_i^{j,l}\delta{X}_i^{j,l} + \delta{C}_i^{j,l}\delta{X}_i^{j,l}\right).
\end{align}
Equation~\eqref{eq:deltaX_corr^TTL} depicts the remaining noise following the subtraction of TTL noise. The right-hand side reveals three contributing factors: the first being the estimation error in the coefficients, the second originates from the sensing noise associated with the DWS, and the third arises from the interplay between estimation error and the DWS readout noise. The readout noise of the DWS is comparatively small, particularly after amplification, with the predominant source stemming from errors in estimating the coupling coefficients in the initial term.
 
Given that the coefficients originate from null TDI combinations, it is vital to ensure that the TTL residual noise within the X combination remains below 0.3 pm/Hz$^{1/2}$ in order to validate the effectiveness of the estimation method. Then we are able to construct other TDI variables for GW extraction that are free from TTL noise.


\section{Simulation setup} \label{sec: setup}
This section details our numerical simulation setup, aiming to validate the efficiency of null TDI channel combination in estimating TTL coefficients and implementing closed-loop control simulation to produce jitter noise.

\subsection{TDI setup}
Utilizing TianQin's orbit as described in Refs.\cite{ye2021eclipse}, which spans a 1 day period, all our simulations employ a 5 Hz sampling rate. Noise levels are quantified using the amplitude spectral density (ASD), defined as the square root of the power spectral density (PSD). The laser frequency noise's ASD is specified by
\begin{equation}
	S_{\nu}^{1/2}(f) = 30\ \text{Hz/Hz}^{1/2}\times\text{NSF}(f)
\end{equation}
where the noise shape function is expressed as
\begin{equation}
	\text{NSF}(f) = \sqrt{1+\left(\frac{4\ \text{mHz}}{f}\right)^4}.
\end{equation}
The biases in TDI delays employed for laser noise reduction are approximately 3 ns.

The fractional frequency noise ASDs of USO and sideband modulation are given by
\begin{equation}
	S_{\text{u}}^{1/2}(f) = 7.0\times10^{-14}\text{/Hz}^{1/2}\times\sqrt{\frac{1\ \text{Hz}}{f}}
\end{equation}
and
\begin{equation}
	S_{\text{m}}^{1/2}(f) = 2.7\times10^{-14}\text{/Hz}^{1/2}\times\text{NSF}(f).
\end{equation}

Polarization and stray light in interferometers are modeled as equivalent displacement noise characterized by an ASD of
\begin{equation}
	S_{\text{opt}}^{1/2}(f) = 0.2\ \text{pm/Hz}^{1/2}\times\text{NSF}(f).
\end{equation}

The various readout noise levels in carrier phase measurements are presented in PSD terms as follows:
\begin{subequations}
	\begin{align}
		S_{\text{ro}}^{\text{isi}}(f) &= (0.50\ \text{pm/Hz}^{1/2}\times\text{NSF}(f))^2, \\
		S_{\text{ro}}^{\text{tmi}}(f) &= (0.15\ \text{pm/Hz}^{1/2}\times\text{NSF}(f))^2, \\
		S_{\text{ro}}^{\text{rfi}}(f) &= (0.15\ \text{pm/Hz}^{1/2}\times\text{NSF}(f))^2.
	\end{align}
\end{subequations}
Similarly, we have
\begin{subequations}
	\begin{align}
		S_{\text{ro,sb}}^{\text{isi}}(f) &= (7.15\ \text{pm/Hz}^{1/2}\times\text{NSF}(f))^2, \\
		S_{\text{ro,sb}}^{\text{tmi}}(f) &= (2.20\ \text{pm/Hz}^{1/2}\times\text{NSF}(f))^2, \\
		S_{\text{ro,sb}}^{\text{rfi}}(f) &= (2.20\ \text{pm/Hz}^{1/2}\times\text{NSF}(f))^2
	\end{align}
\end{subequations}
for sideband measurements.

The simulation also incorporates the acceleration noise of the test mass, featuring an ASD of
\begin{equation}
	S_{\text{TM}}^{1/2}(f) = 1.0\ \text{fm/s}^2\text{/Hz}^{1/2}\times\sqrt{1+\left(\frac{0.1\ \text{mHz}}{f}\right)^2}.
\end{equation}

Concerning the jitter noise that adds to the TTL noise, the angular jitters of both the satellite and the MOSA are generated as outputs from the closed-loop control simulation. Furthermore, the simulation will integrate the real DWS noise. When both simulation sets use identical physical quantities, the magnitude of their noise shape functions will be precisely the same.

\subsection{Closed-loop control setup}
Within the science mode, the DFPCS utilizes three actuator types and four sensor categories. The actuators include micro-thrusters, inertial sensors, and MOSA drive mechanisms. The sensor types are as follows: a). Interferometric displacement sensors (IFO) on the satellite measure relative TM displacement along the sensitive axis in the $\mathcal{OA}$ frame and TM's 2-DoF rotation relative to the $\mathcal{OA}$ frame (pitch and yaw). b). Inertial sensors provide 6-DoF TM position/attitude from capacitive measurements. c). DWS sensors determine pitch and yaw angles between the telescope's optical axis and incoming laser beams. Performance details for sensors and actuators are presented in Table \ref{TableClosedSetup}. Note that the angular jitter sensing noise in the table must be divided by $M = 300$, which is the telescope's and imaging system's total magnification on OB.
\begin{table}[ht]
\caption{Sensing and actuation noise settings used in the DFPC simulations. All noises listed are assumed white within the TianQin mission bandwidth.}
\label{TableClosedSetup}
\centering
\begin{ruledtabular}
\begin{tabular}{lll}
Source & Noise Type & Noise PSD \\
\colrule
Micro-thruster & Force & $1 \times 10^{-7}~~\mathrm{N/\sqrt{Hz}}$ \\
Inertial sensor & Force & $1 \times 10^{-14}~\mathrm{N/\sqrt{Hz}}$ \\
    & Torque & $1 \times 10^{-15}~\mathrm{nm/\sqrt{Hz}}$ \\
    & Position sensing & $1 \times 10^{-9}~~\mathrm{m/\sqrt{Hz}}$ \\
    & Attitude sensing & $1 \times 10^{-7}~~\mathrm{rad/\sqrt{Hz}}$ \\
MOSA & Torque & $1 \times 10^{-9}~~\mathrm{nm/\sqrt{Hz}}$ \\
IFO  & Position sensing & $1 \times 10^{-12}~\mathrm{m/\sqrt{Hz}}$ \\
     & Attitude sensing & $1 \times 10^{-9}~~\mathrm{rad/\sqrt{Hz}}$ \\
DWS  & Attitude sensing & $10 \times 10^{-9}~\mathrm{rad/\sqrt{Hz}}$ \\
\end{tabular}
\end{ruledtabular}
\end{table}


\section{Result}\label{sec: result}
The focus of this research is the analysis of TTL noise subtraction with realistic jitters simulated from closed-loop dynamics. In this section, the simulated closed-loop jitters will be employed to produce TTL noise in the TDI simulations. Meanwhile, the DWS readouts will be used to estimate the TTL coupling coefficient, which can then be used to subtract the TTL noise from the intermediary variable $\eta_i$.

\subsection{Simulated jitter}\label{sec: jitterResult}
\begin{figure}[ht]
\begin{minipage}{0.48\textwidth}
  \centering
  \includegraphics[width=\textwidth]{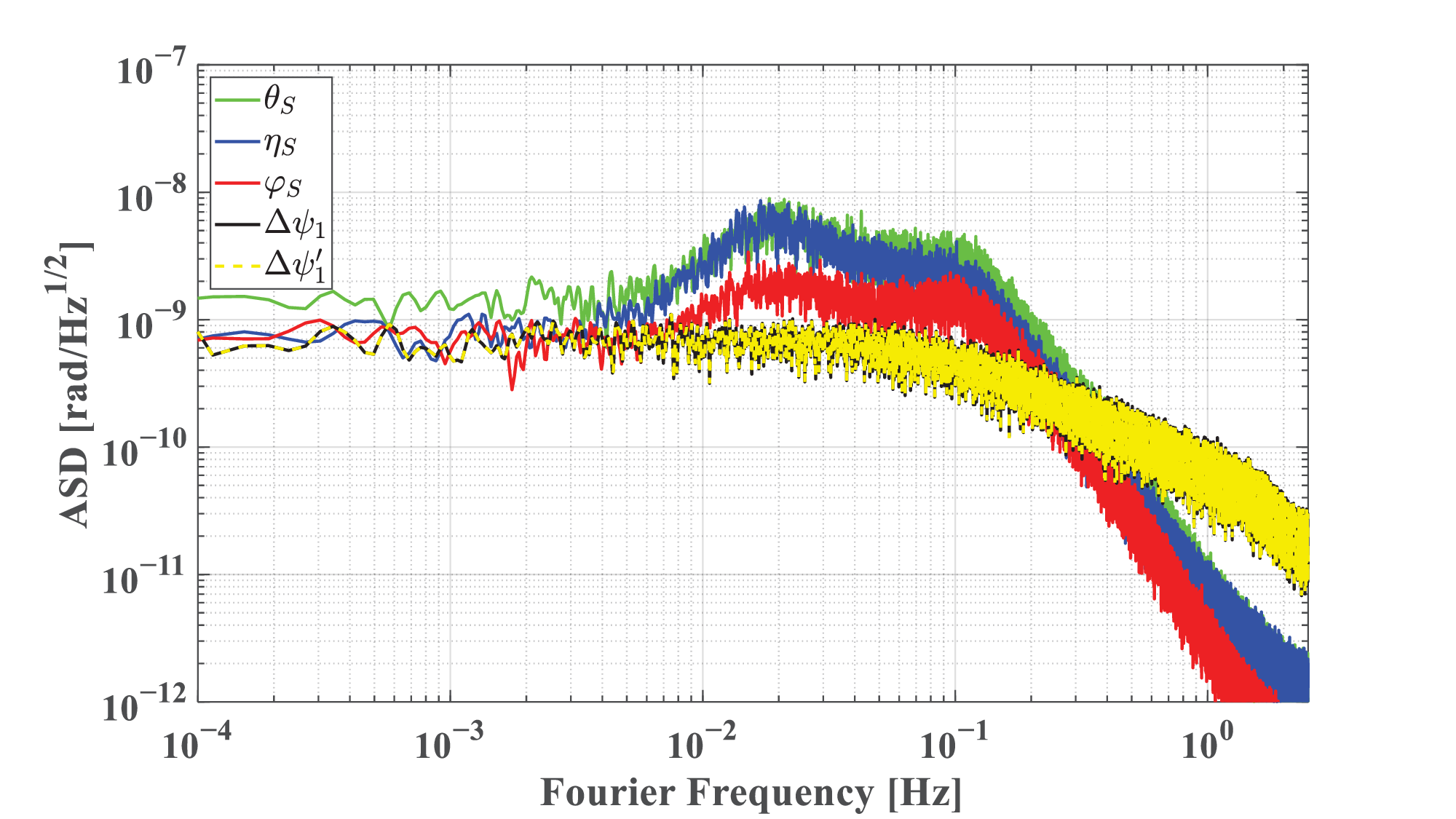}
\end{minipage}
\begin{minipage}{0.48\textwidth}
  \centering
  \includegraphics[width=\textwidth]{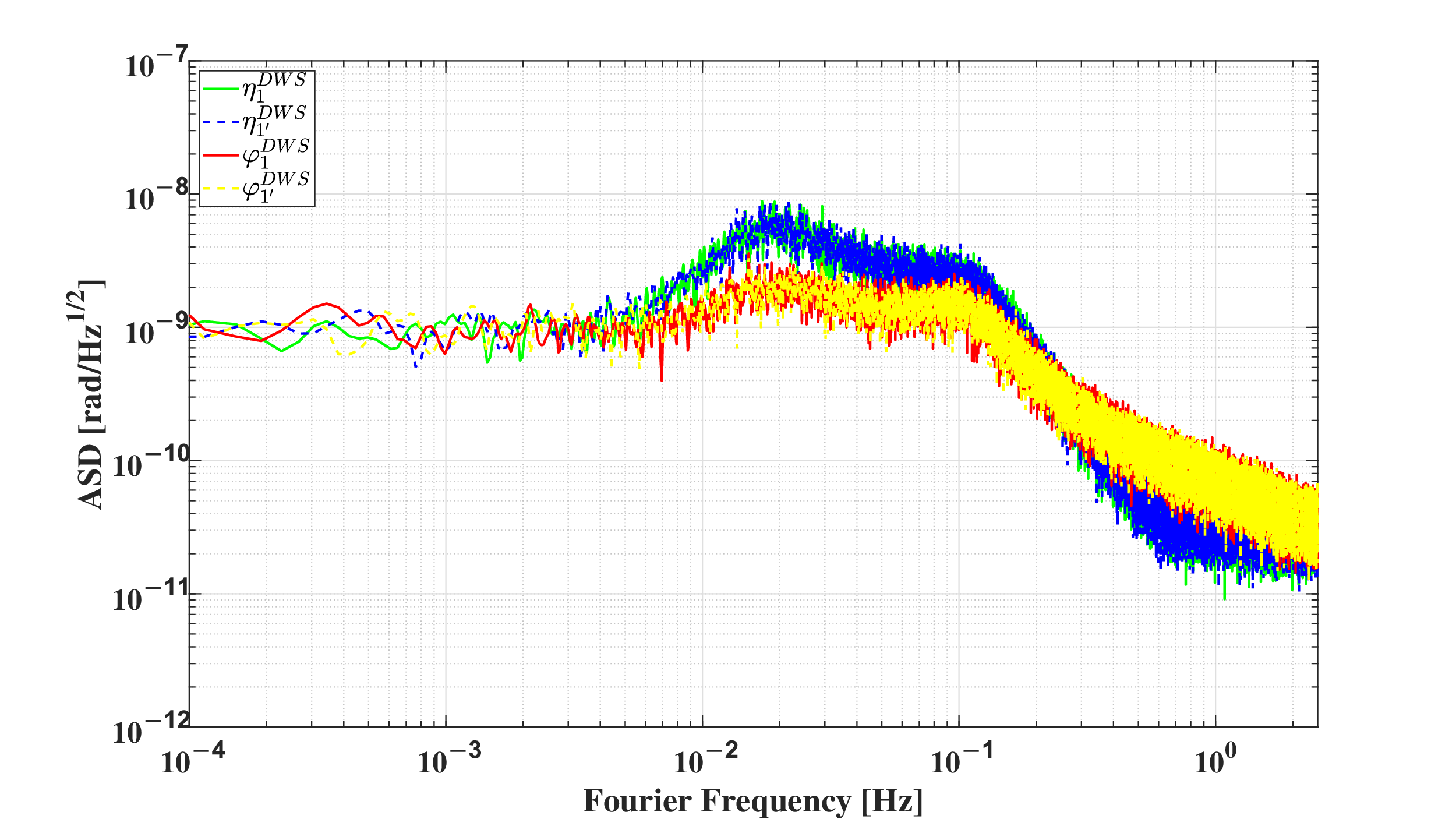}
\end{minipage}
\caption{\label{fig:Jitter_double} The ASD plots of the satellite’s three-axis jitter and MOSA yaw angle jitter under symmetric rotation mode (the upper plot), along with the two DWS readout signals (the lower plot). The true jitter data are multiplied by the coupling coefficients to simulate TTL noise in the TDI. The DWS readout data will subsequently be utilized for TTL noise estimation. }
\end{figure}

\begin{figure}[ht]
\begin{minipage}{0.48\textwidth}
  \centering
  \includegraphics[width=\textwidth]{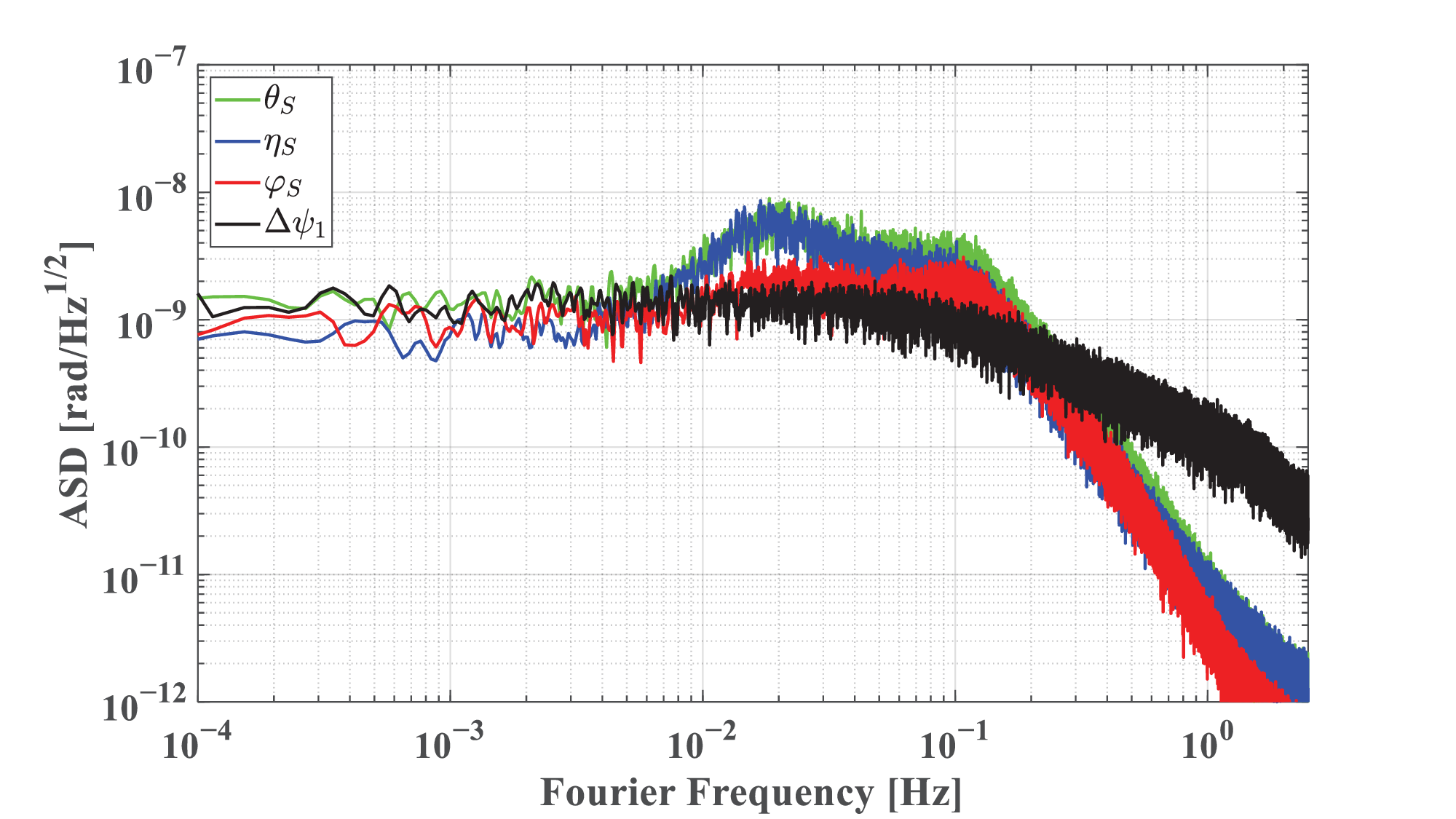}
\end{minipage}
\begin{minipage}{0.48\textwidth}
  \centering
  \includegraphics[width=\textwidth]{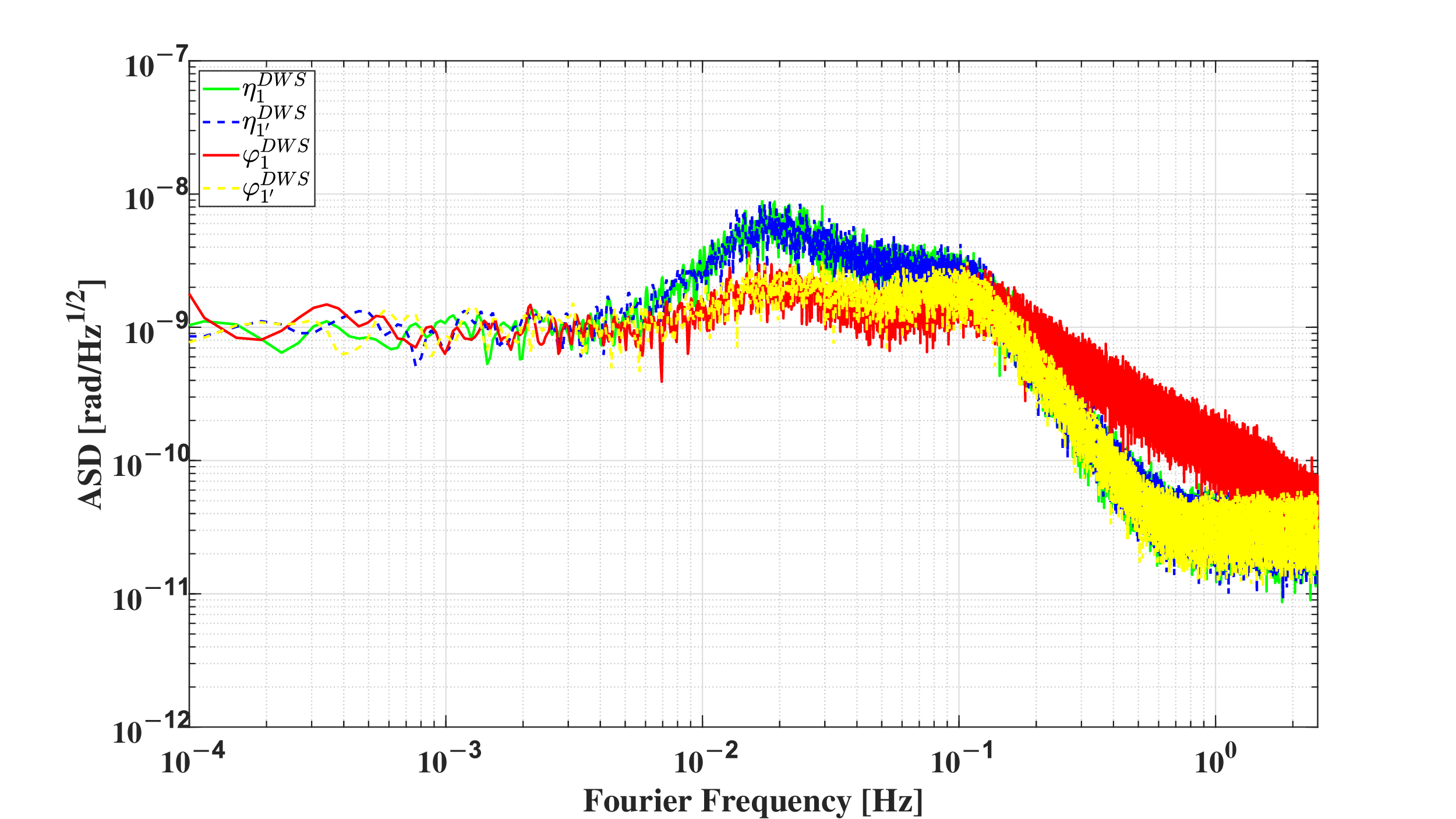}
\end{minipage}
\caption{\label{fig:Jitter_single} The ASD plots of the satellite’s three-axis jitter and MOSA yaw angle jitter under single MOSA rotation mode (the upper plot), along with the two DWS readout signals (the lower plot). }
\end{figure}

For the jitter measurements simulated from the closed-loop dynamics, we examine two scenarios below. Fig.~\ref{fig:Jitter_double} shows that the closed-loop dynamics jitters of the satellite and MOSAs, which we use to generate the TTL noise, and the lower subplot shows the DWS readouts that are used to estimate the coefficients. Second, Fig.~\ref{fig:Jitter_single} shows the single MOSA rotation mode, where only MOSA$_1$ rotates for compensating the breathing angle. From the simulation, the pitch pointing performance of these two modes both satisfies the pointing stability requirement of 10 nrad/Hz$^{1/2}$.

Within the frequency range of 0.01–0.1 Hz, micro-thruster noise increases jitter across all three satellite axes. This increase leads to significant cross-correlation in satellite jitter projections along both MOSA directions, potentially hindering TTL estimation. Within the 0.1–1 Hz frequency range, the figures both indicate that the yaw pointing jitters in DWS readouts are primarily influenced by the jitter of the MOSAs. This phenomenon arises from the significant discrepancy in moments of inertia, where the satellite exhibits a moment of inertia (e.g., diag [583 kg$\cdot$m$^2$, 583 kg$\cdot$m$^2$, 1125 kg$\cdot$m$^2$]), which is considerably larger compared to the MOSAs (e.g., diag [1.51 kg$\cdot$m$^2$, 1.55 kg$\cdot$m$^2$, 1.55 kg$\cdot$m$^2$]). For the same reason, jitter around the satellite's z-axis is less compared to the other axes. Frequencies such as 10$^{-4}$ to 10$^{-2}$ Hz and above 1 Hz are not ideal for TTL estimation. When analyzing the satellite's three-axis jitter as it is projected onto the MOSA, it is clear that, at high frequencies, the projection reduces, whereas the MOSA jitter tends to be pronounced. In addition to the ASD jitter, it is observed that under the symmetric rotation mode, there is a strong correlation between the jitter magnitudes of the two MOSAs, which is consistent with Eq.~\eqref{eq:symmetric}. Contrarily, this correlation does not manifest in the single-axis rotation mode of the MOSAs.

In summary, given that the single MOSA rotation mode will cause the two MOSAs to have no correlations and the gradual diminution of the satellite's projection at high frequencies, we deduce that utilizing high-frequency data from the single MOSA rotation mode is advantageous for improving coefficient estimation. The forthcoming Sec.~\ref{sec: subtratcionResult} will exhibit these results.

\subsection{Subtraction result}\label{sec: subtratcionResult}
\begin{figure}[ht]
\centering
\includegraphics[width=0.48\textwidth]{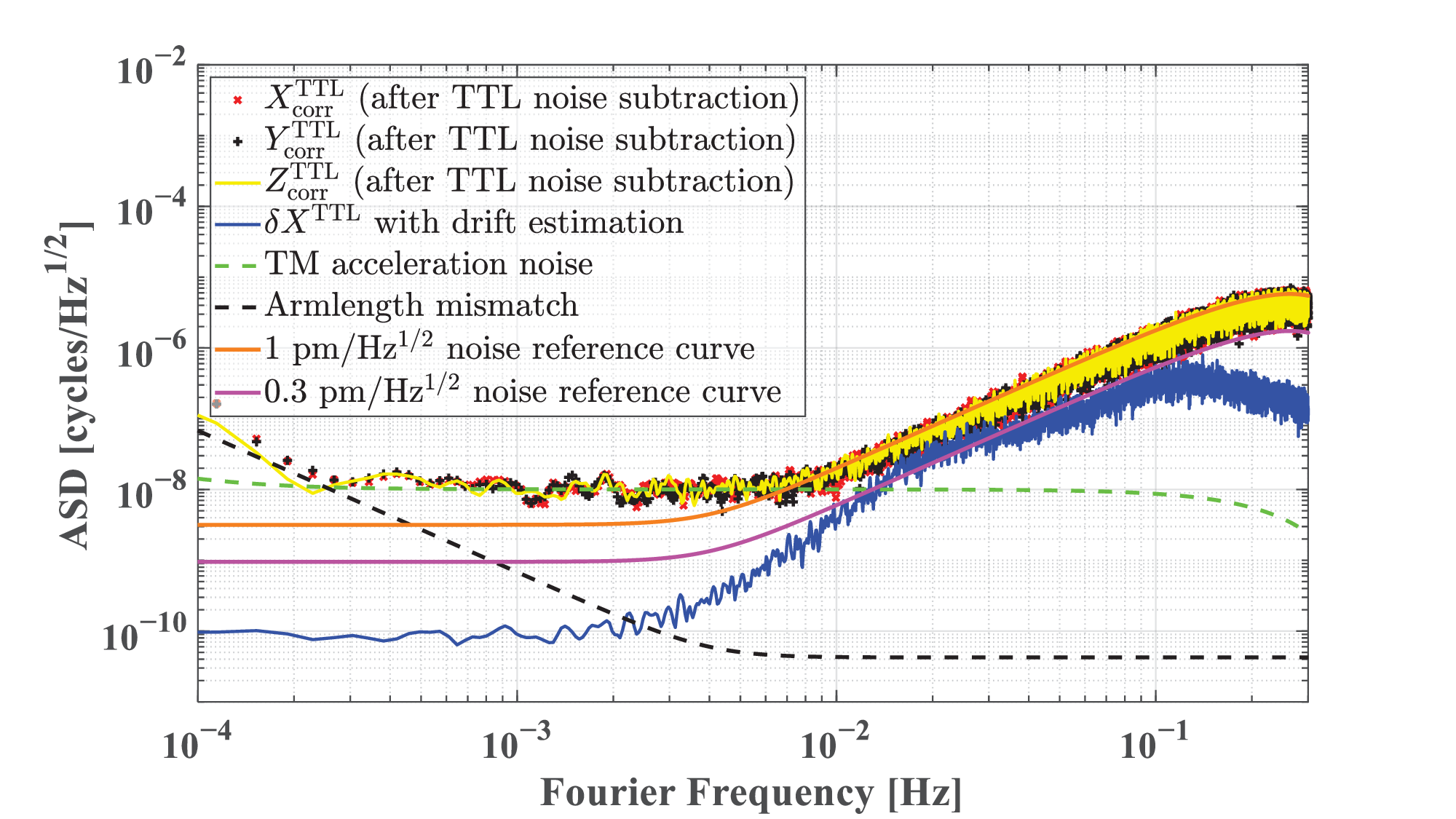}
\caption{\label{fig:doubleRes} Residual noise level in X, Y, and Z after subtracting the TTL noise and clock noise with closed-loop dynamics under symmetric MOSA rotation mode. The blue traces represent TTL noise residuals in X channel, it is noticed that in this situation, approach does not fully meet the requirements for subtraction. }
\end{figure}

In the symmetric MOSA rotation mode, there is a significant correlation between the jitters of the two MOSAs. Estimating coefficients across various frequency bands, such as 0.01–0.1 Hz and the higher 0.1–1 Hz band, both result in residual noise that partially exceeds the 0.3 pm/Hz$^{1/2}$ requirement. For clarity, we only present the results for the 0.1–1 Hz range in Fig.~\ref{fig:doubleRes}. This outcome proves unfavorable for the mission. Thus, the subsequent discussion in \ref{sec: discussion} will explore methods to enhance estimation accuracy under these conditions.

\begin{figure}[ht]
\begin{minipage}{0.48\textwidth}
  \centering
  \includegraphics[width=\textwidth]{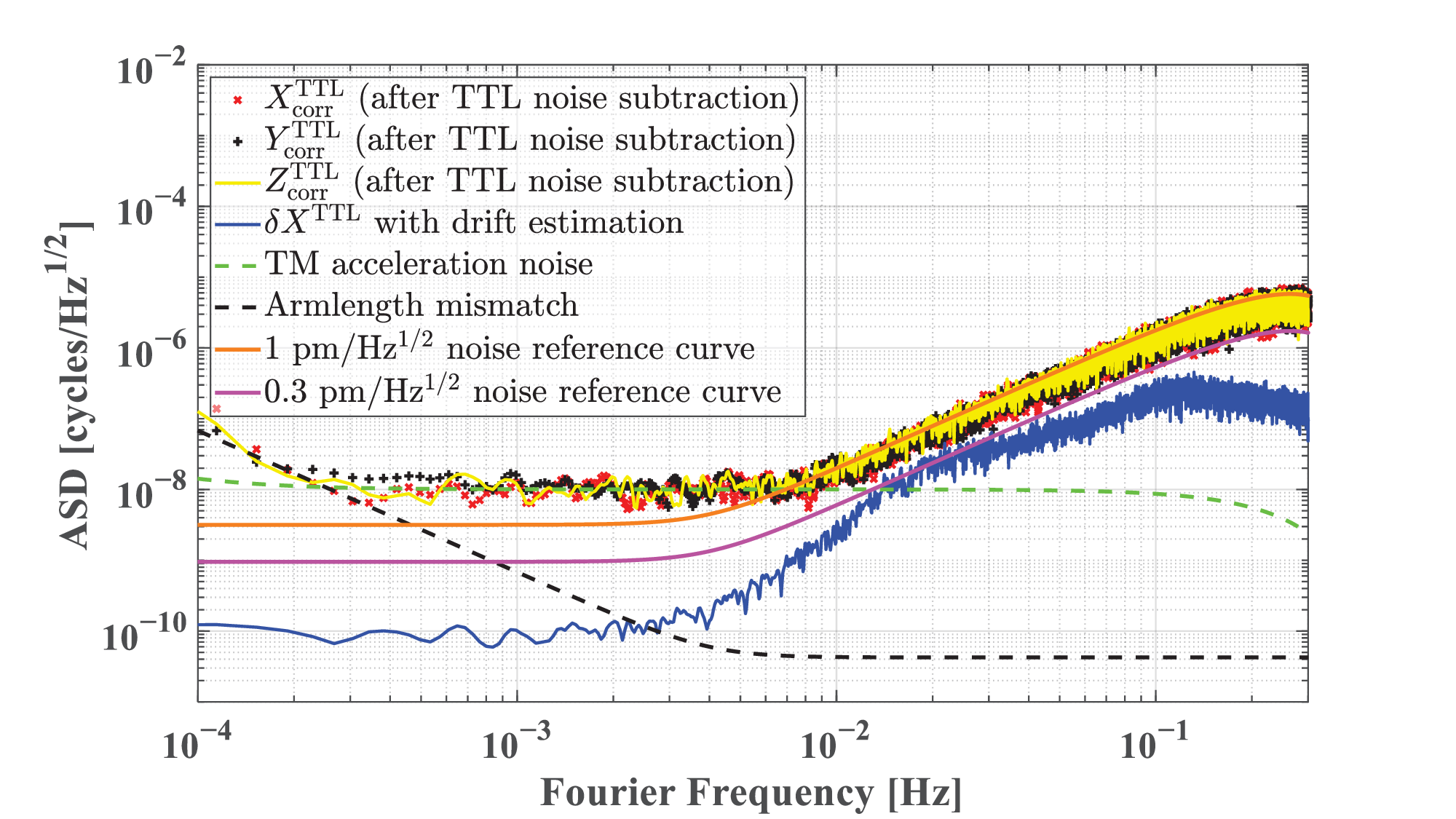}
\end{minipage}
\begin{minipage}{0.48\textwidth}
  \centering
  \includegraphics[width=\textwidth]{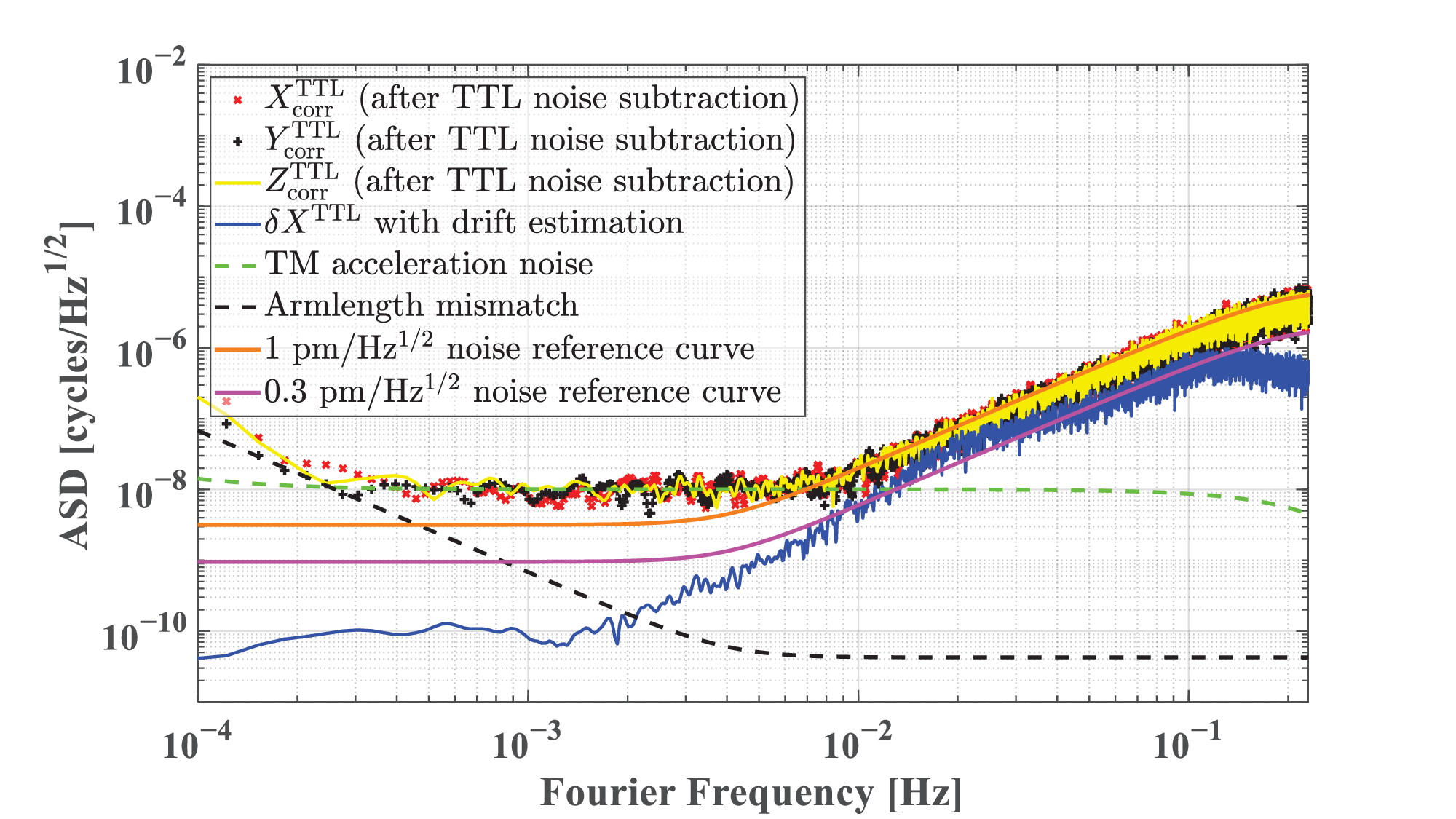}
\end{minipage}
\caption{\label{fig:singleRes} Residual noise level under single MOSA rotation mode. The upper plots illustrates the subtraction performance achieved using coefficients estimated from high-frequency data, while the lower figure demonstrates the corresponding result obtained using coefficients derived from low-frequency data. }
\end{figure}

Under single MOSA rotation mode, we calculate two estimation subtraction results using different frequency band data for comparison. For utilizing high-frequency data (0.1–1 Hz) for TTL coupling coefficient estimation, the result in the upper subplot of Fig.~\ref{fig:singleRes} proves particularly effective and the residual noises will be proved to be under 0.3 pm/Hz$^{1/2}$. For the data in the lower frequency (0.01-0.1 Hz), the subtraction performance will be relatively worse, especially around the 0.02 Hz, where the micro-thruster noise will be dominant.  

In addition, to better study the influence of DWS noise on the estimation of coefficients using the least squares method, we introduced an additional bias term to the measurement noise. The results remain unaffected, primarily because an overall biases increase in noise does not influence the estimation of the linear slope. Therefore, the accuracy of the estimated coefficients is not altered by the measurement bias in the DWS signal.

In summary, reducing the correlation between jitters of the two MOSAs in estimation improves the least squares estimation, resulting in more accurate TTL coefficients estimation. Our analysis demonstrates that adopting the single MOSA rotation mode during scientific observation periods, and using the jitter data from the 0.1-1 Hz frequency band, provides enhanced TTL noise suppression. This meets the strict criterion of minimizing residual TTL noise below the 0.3 pm/Hz$^{1/2}$ threshold.

\section{Discussion}\label{sec: discussion}
The previous Sec.~\ref{sec: subtratcionResult} indicates that symmetric MOSA rotation mode does not sufficiently reduce residual noise to the desired level. To improve coefficient estimation accuracy, we explore two optimization strategies and use simulations to demonstrate them. One approach involves employing various null TDI combinations, while the other focuses on modulating artificial MOSA maneuvers. Both methods will present performance evaluations demonstrating better noise suppression while maintaining operational feasibility.

\subsection{Different Null channel combinations}
The key point of improving estimation accuracy is trying to reduce the matrix ill-conditioning during the least squares estimation process. The combinations we typically employ, i.e., null TDI channel combination $C_3^{12}$, are invariably identical, meaning the two variables from the same satellite always appear in pairs. This leads to inaccuracies in our estimations, necessitating the examination of additional combinations.

In particular, because the yaw jitter characteristics of the two MOSAs are strongly correlated under symmetric rotation mode, the adoption of the $C_3^{12}$ combination will lead to increased errors in coupling coefficient determination. From the examination in Appendix \ref{sec: AppendixB} and the illustrative Fig.~\ref{fig:C312C314Demo}, it is evident that in the frequently employed $C_3^{12}$ configuration, the noise introduced by jitter in the interferometric signals results in similar impacts on the pitch and yaw angles of the MOSAs. Additionally, the time delays for $i$ and $i^\prime$ are identical, which causes an intrinsic inseparability between the two MOSAs located on the same satellite, as outlined in Eq.~\eqref{eq:C312Full}. However, in the $C_3^{14}$ combination (see Eq.~\eqref{eq:C314Full}), the time delays of signals $i$ and $i^\prime$ will have some differences, which is supportive for the coefficient estimation.
\begin{figure}[ht]
\begin{minipage}{0.48\textwidth}
  \centering
  \includegraphics[width=\textwidth]{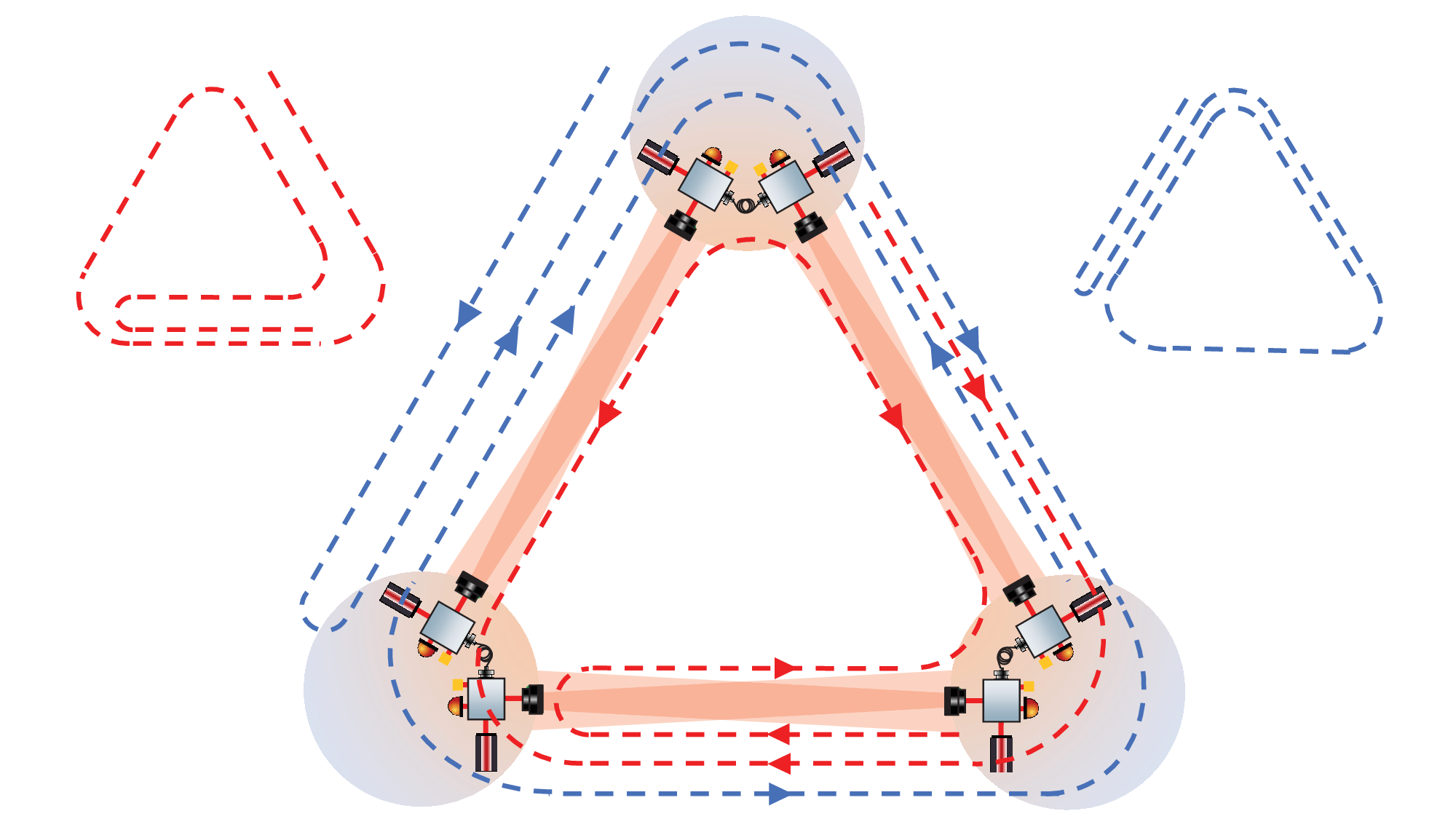}
\end{minipage}
\begin{minipage}{0.48\textwidth}
  \centering
  \includegraphics[width=\textwidth]{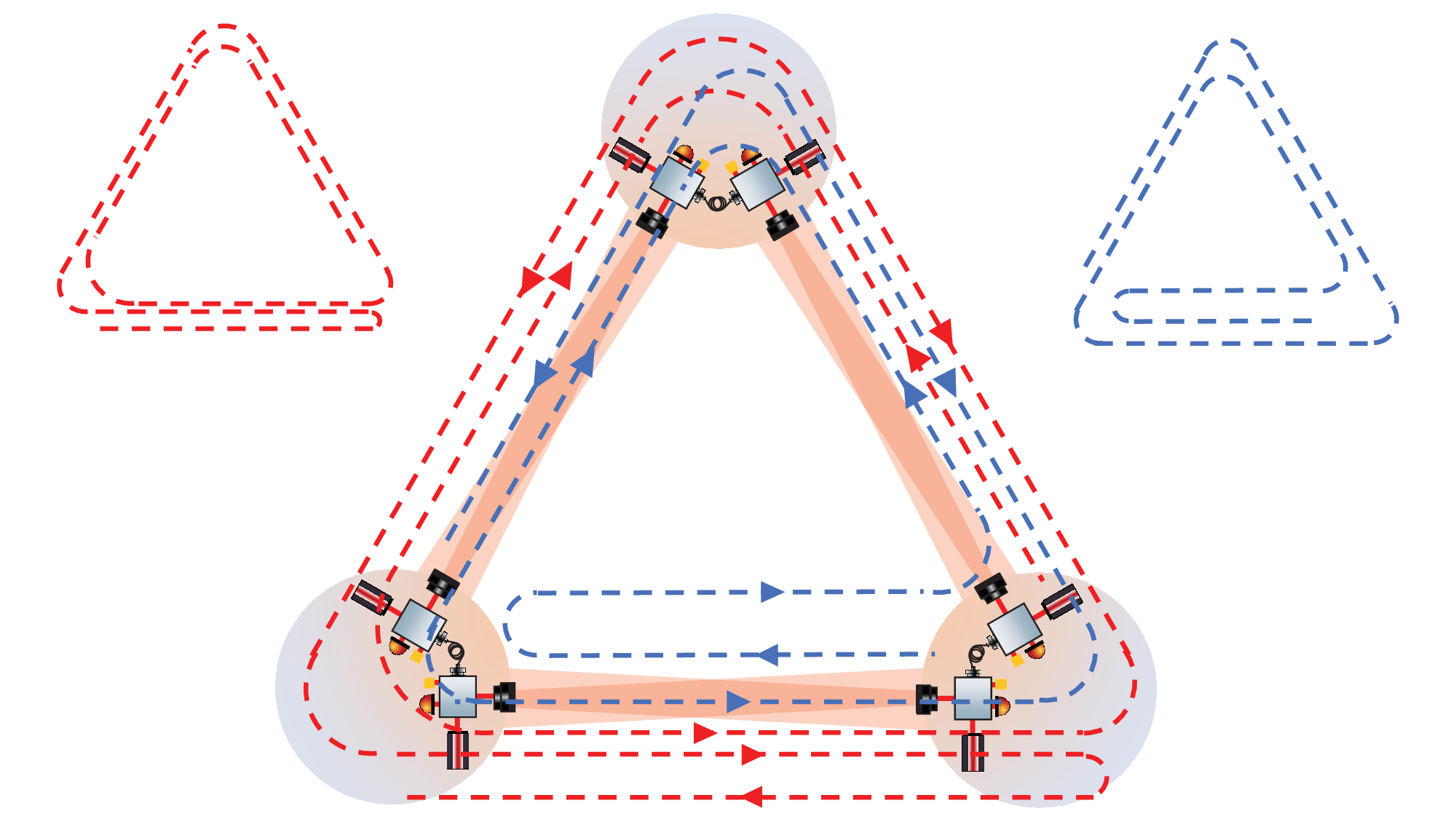}
\end{minipage}
\caption{\label{fig:C312C314Demo} Illustration of a virtual optical pathway synthesized from two null channel configurations, designated as $C_3^{12}$ (illustrated in the upper plot) and $C_3^{14}$ (depicted in the lower plot). In the $C_3^{14}$, there exist six connections between $i$ and $j'$, whereas $i'$ is connected to $k$ by four links, as is $j$ to $k'$. Conversely, the $C_3^{12}$ arrangement features a uniform distribution of four connections for each pair of satellites, promoting symmetry at the expense of reducing estimation efficiency. }
\end{figure}

\begin{figure}[ht]
\begin{minipage}{0.48\textwidth}
  \centering
  \includegraphics[width=\textwidth]{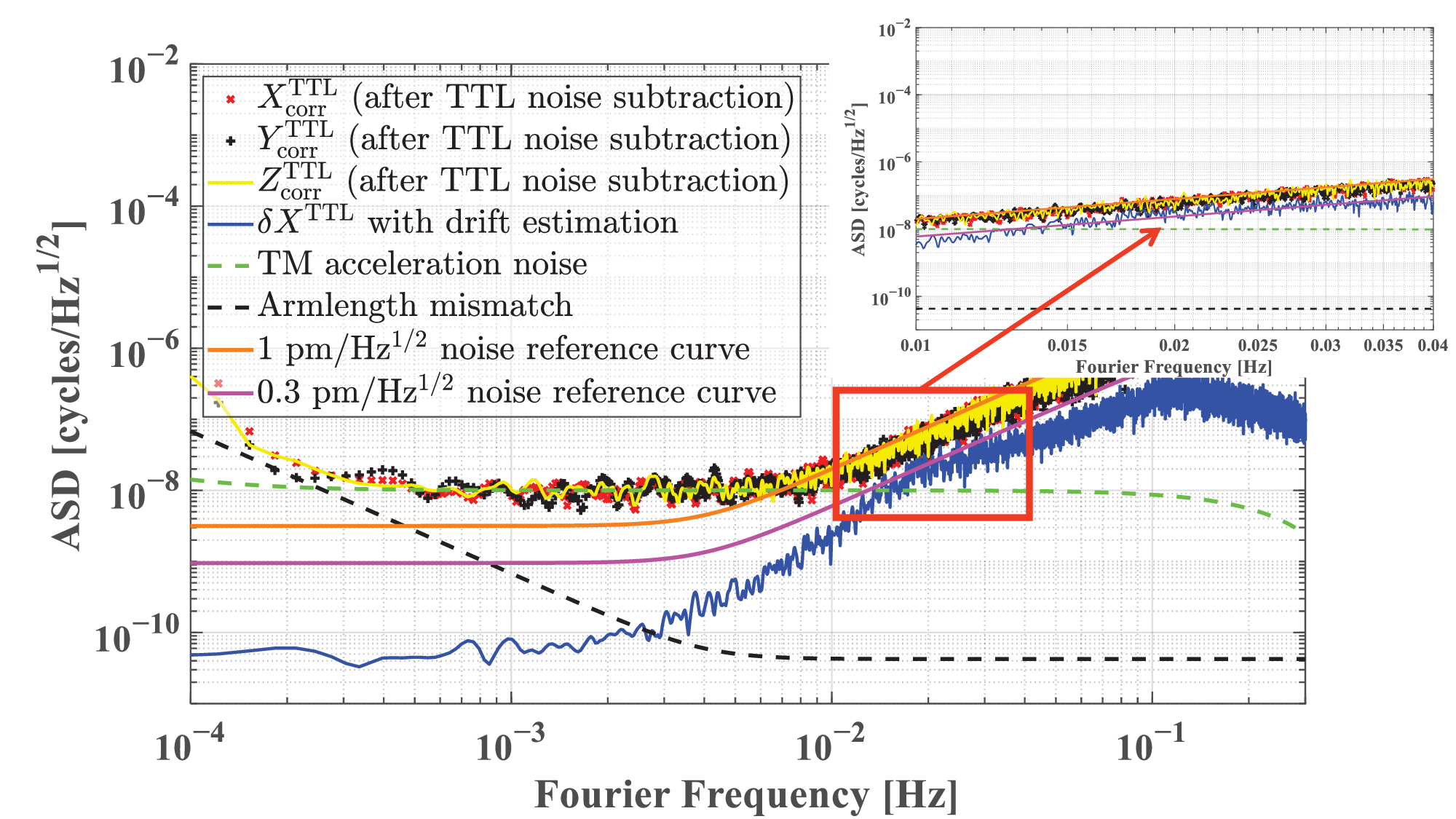}
\end{minipage}
\begin{minipage}{0.48\textwidth}
  \centering
  \includegraphics[width=\textwidth]{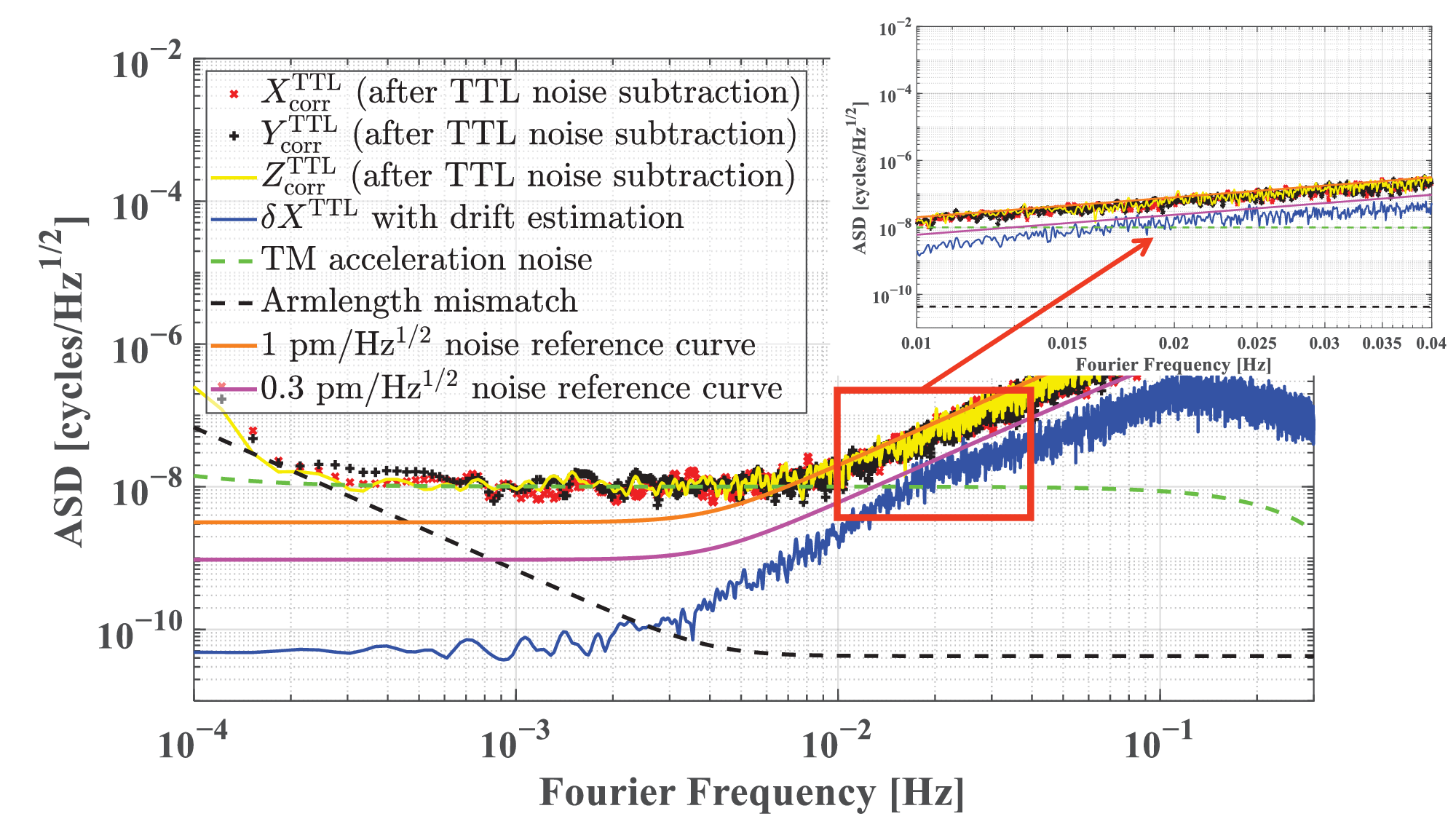}
\end{minipage}
\caption{\label{fig:C312C314} The level of residual noise observed subsequent to the application of $C_3^{12}$ (depicted in the upper graph) or $C_3^{14}$ (illustrated in the lower graph) for the mitigation of TTL noise. Under similar conditions, it is clear that simply changing the TDI combination can improve estimation accuracy. Moreover, using the $C_3^{14}$ combination facilitates the reduction of noise to levels beneath the recognized benchmark. }
\end{figure}

To test this hypothesis, we utilized combinations of $C_3^{14}$ and $C_3^{14}$ to analyze their subtraction outcomes. By systematically rotating the reference satellite, we calculated three separate sets of estimated coupling coefficients and subsequently performed an inverse-variance-weighted average on these datasets. As shown in Fig.~\ref{fig:C312C314}, the upper subplot corresponds to the $C_3^{12}$ combination, and this method does not produce any marked improvement in outcomes. In contrast, when applying the $C_3^{14}$ combination, as demonstrated in the lower subplot, we successfully suppressed the symmetric rotation-induced noise below the reference threshold 0.3 pm/Hz$^{1/2}$. The outcome guarantees that, within a symmetric rotation mode, employing the $C_3^{14}$ for the least squares estimation of TTL coupling coefficients can be executed with enhanced precision and reliability.

The challenge of not fulfilling the requirement occurs because of the closed-loop dynamics inherent in the symmetrically rotating MOSA. This can now be tackled through the application of different null TDI channel combinations, concentrating exclusively on the algorithmic strategy. We further evaluate the null Time Delay Interferometry (TDI) combinations, such as $C_{26}^{16}$, $C_{27}^{16}$, $C_{28}^{16}$, as discussed in Ref.~\cite{hartwig2022characterization}. The result of the subtraction stays nearly consistent, aligning with the findings reported in Ref.\cite{chen2025characteristics}. For clarity, we propose using the $C_3^{14}$ for the estimation procedure. Thus, if the $C_3^{12}$ is inadequate in achieving the desired residual noise level during TTL noise subtraction, we recommend exploring alternative configurations, such as $C_3^{14}$.

\subsection{MOSA maneuver}
The impact of maneuvers on satellites has been examined in Refs.\cite{wegener2025design, houba2022lisa, houba2022optimal} with the aim of optimizing TTL coefficient estimation. However, the potential improvement in coefficient estimation through the introduction of an artificial signal at the marginal frequency, such as 1 Hz, while simultaneously meeting DFPC requirements during the science mode, warrants further investigation. 

\begin{figure}[ht]
\centering
\includegraphics[width=0.48\textwidth]{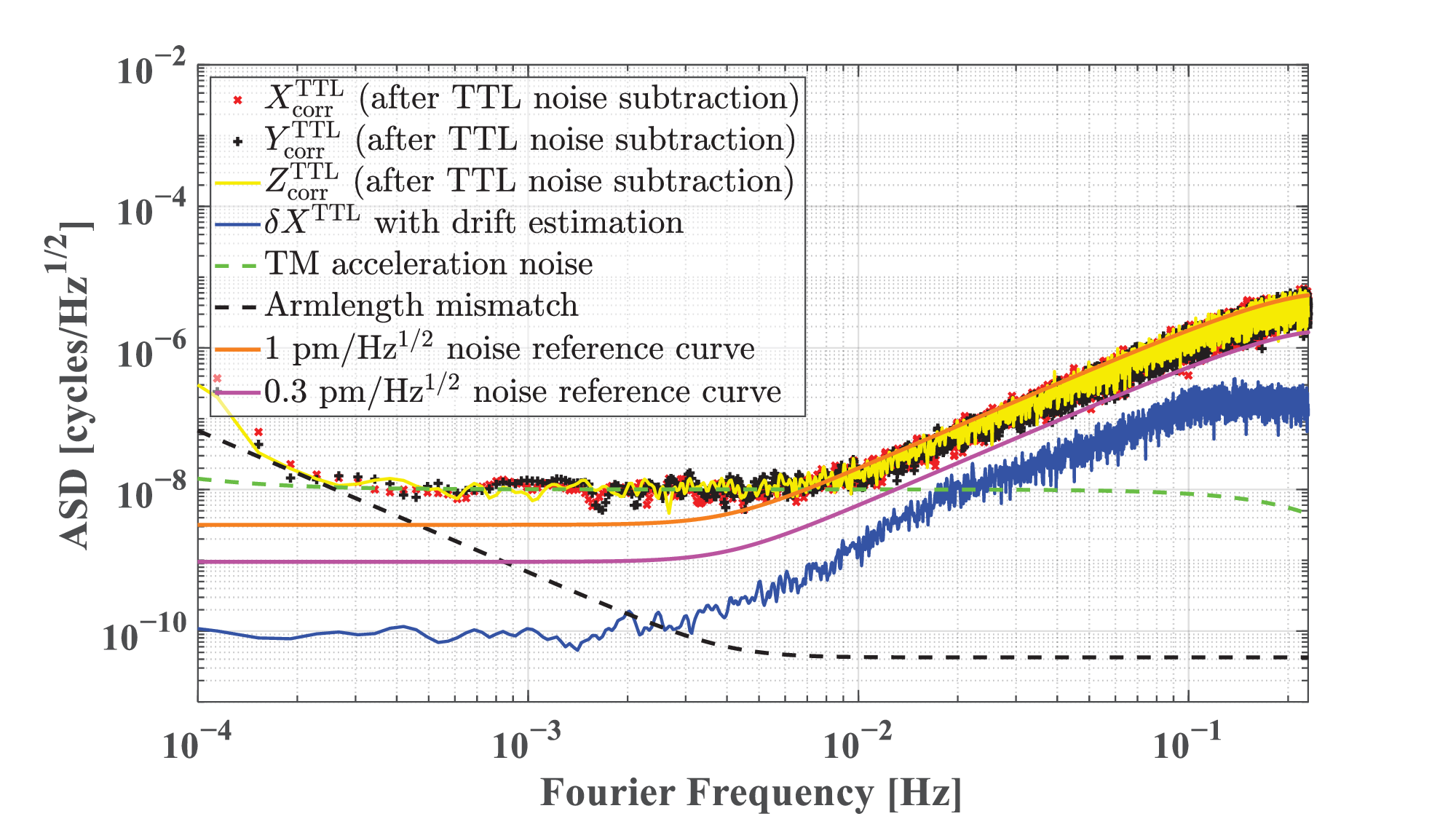}
\caption{\label{fig:Result_modulation} Residual noise level after modulate MOSA jitter. Modulating the MOSAs' jitters offer substantial benefits for coefficient calibration. However, the introduction of an additional jitter signal may potentially introduce other noises, necessitating further investigation. }
\end{figure}

For the modulated jitter, to meet the requirements of drag-free pointing control, we select a modulation jitter of 0.5 nrad at a single frequency of 1 Hz. If the frequency were higher, the tracking capability of the control system would deteriorate, and the inherent jitter of the MOSA would become more pronounced. Consequently, a sinusoidal modulation of 0.5 nrad at 1 Hz, each with a distinct initial phase, was applied to the six MOSAs across the three satellites to facilitate drag-free controller tracking. The simulation results of the jitter continue to meet the DFPC performance requirements. We incorporated the simulated jitters data into the null channel process analysis. As can be observed from the Fig.~\ref{fig:Result_modulation}, after the modulation jitters are applied, the original method effectively suppresses the noise to below 0.3 pm/Hz$^{1/2}$.

Although this method has proven feasible in DFPC, the noise induced by the modulated jitter may introduce extra risks to mission detection. Consequently, it remains a backup option for the TianQin project.


\section{Conclusion}\label{sec: conclusion}
Using a comprehensive closed-loop dynamics model, our simulations show that the satellites and MOSAs angular jitters can introduce strong correlations in the time domain that diminish the TTL coefficient estimation accuracy in the null TDI combination. To mitigate this problem, selecting an appropriate frequency range informed by the characteristics of the jitter feature is crucial for decreasing the impact and improving the accuracy of estimations. Based on that, we evaluated two kinds of pointing control schemes, the single MOSA rotation mode and the symmetric MOSA rotation mode. The simulation results indicate that the single MOSA rotation mode enhances TTL estimation accuracy when compared to the symmetric rotation mode, especially at high frequencies. In summary, in TianQin, utilizing high frequency data under the single MOSA rotation mode for estimating the TTL coefficients via the null TDI channel can effectively reduce the residual TTL noise level to under 0.3 pm/Hz$^{1/2}$.

In order to enhance the estimation accuracy, we discussed two methods in Sec.~\ref{sec: discussion}. To begin with, we identified that employing different null TDI channel combinations, such as $C_3^{14}$, can enhance estimation accuracy. The reason for this is that it creates different combinations of time delays between two sequences of jitter signals that come from the same satellite. Furthermore, to differentiate the jitter between the two MOSAs, artificial jitter modulation was introduced to each of the six MOSAs within the constellation. This notably enhances estimation accuracy, particularly when a correlation exists between the jitter of the two MOSAs under the symmetric MOSA rotation mode. In addition, closed-loop simulations demonstrated the effectiveness of this modulation at a modulation amplitude of 0.5 nrad at a frequency of 1 Hz. Notably, although modulation jitter applied to both MOSAs has been demonstrated as feasible through DFPC simulation, greater caution must be exercised during the scientific observation period. 

Considering our utilized null channel combinations and the derived coefficients from high-frequency band data, coupled with the findings of \cite{hartig2025postprocessing} indicating an negligible impact of GW on TTL subtraction, GW signals have therefore been omitted from the current simulations. Future studies should include GW responses to enable thorough TTL mitigation analyses across diverse TDI combinations and frequency bands. Furthermore, as research on DWS is still in progress, we have substituted its contribution with basic white noise, omitting its nonlinear characteristics from this paper. A comprehensive investigation of these elements will be essential in subsequent work.


\begin{acknowledgments}
The authors thank Zhizhao Wang for providing the basic program and for his valuable discussions and feedback. X. Z. is supported by the National Key R\&D Program of China (Grant No. 2022YFC2204600) and NSFC (Grant No. 12373116).
\end{acknowledgments}


\appendix

\section{Small-angle approximation}\label{sec: AppendixA}
This appendix provides an in-depth examination of the transformation matrix associated with the jitters of MOSAs and satellites. These jitters, being minor angles of coordinate rotation, allow the matrix to be expressed using the small-angle approximation, i.e.
\begin{widetext}
\begin{equation}
\mathcal{M}(\theta, \eta, \varphi) = \mathcal{R}_z(\varphi)\mathcal{R}_y(\eta)\mathcal{R}_x(\theta)   
= 
\begin{bmatrix}
\cos(\varphi) & -\sin(\varphi) & 0 \\
\sin(\varphi) & \cos(\varphi) & 0 \\
0 & 0 & 1
\end{bmatrix}
\begin{bmatrix}
\cos(\eta) & 0 & \sin(\eta) \\
0 & 1 & 0 \\
-\sin(\eta) & 0 & \cos(\eta)
\end{bmatrix}
\begin{bmatrix}
1 & 0 & 0 \\
0 & \cos(\theta) & -\sin(\theta) \\
0 & \sin(\theta) & \cos(\theta)
\end{bmatrix},
\end{equation}

\begin{equation}
\mathcal{M}(\theta, \eta, \varphi) =
\begin{bmatrix}
\cos(\varphi)\cos(\eta) & -\sin(\varphi)\cos(\theta) + \cos(\varphi)\sin(\eta)\sin(\theta) & \sin(\varphi)\sin(\theta) + \cos(\varphi)\sin(\eta)\cos(\theta) \\
\sin(\varphi)\cos(\eta) & \cos(\varphi)\cos(\theta) + \sin(\varphi)\sin(\eta)\sin(\theta) & -\cos(\varphi)\sin(\theta) + \sin(\varphi)\sin(\eta)\cos(\theta) \\
-\sin(\eta) & \cos(\eta)\sin(\theta) & \cos(\eta)\cos(\theta)
\end{bmatrix}
\approx
\begin{bmatrix}
1 & -\varphi & \eta \\
\varphi & 1 & -\theta \\
-\eta & \theta & 1
\end{bmatrix}.
\end{equation}
\end{widetext}
Then for the coordinate transformation matrices, we have
\begin{itemize}
    \item $R_{OA_1^*}^{OA_1}$ is the coordinate transformation matrix from the $\mathcal{OA}_1^*$ frame to the $\mathcal{OA}_1$ frame, which can be written as
    \begin{equation}\label{eq:OA*2OA}
        R_{OA_1^*}^{{OA_1}} = \left[ {
        \begin{array}{*{20}{c}}
        1 & {-{\varphi}_1^\text{DWS}}&{{\eta}_1^\text{DWS}}\\
        {{\varphi}_1^\text{DWS}}&1&{-{\theta}_1^\text{DWS}}\\
        {-{\eta}_1^\text{DWS}}&{{\theta}_1^\text{DWS}}&1
        \end{array}} \right].
    \end{equation}

    \item $R_{OA_1}^S$ is the coordinate transformation matrix from the $\mathcal{OA}_1$ frame to the $\mathcal{S}$ frame, which can be written as
    \begin{equation}\label{eq:OA2S}
        R_{{OA_1}}^S = \left[ {
        \begin{array}{*{20}{c}}
        {\cos({\psi_1})}&{ - \sin({\psi_1})}&0\\
        {\sin({\psi_1})}&{\cos \left( {\psi_1} \right)}&0\\
        0&0&1
        \end{array}} \right] \left[ {\begin{array}{*{20}{c}}
    1&{ - \Delta {\psi_1}}&0\\
    {\Delta {\psi _1}}&1&0\\
    0&0&1
    \end{array}} \right],
    \end{equation}
    where the ${\psi_1}$ is the basis of the opening angle of the MOSA$_1$ due to the breathing angle of the constellation and the $\Delta{\psi_1}$ means the jitter of the MOSA$_1$ relative to the basis.

    \item $R_{OA_1^*}^{S^*}$ is the coordinate transformation matrix from the $\mathcal{OA}_1^*$ frame to the $\mathcal{S}^*$ frame
    \begin{equation}\label{eq:OA*2S*}
    \begin{array}{l}
    R_{OA_1^*}^{S^*} = \left[ {\begin{array}{*{20}{c}}
    {\cos \left( {\psi_1} \right)}&{ - \sin \left( {\psi_1} \right)}&0\\
    {\sin \left( {\psi_1} \right)}&{\cos \left( {\psi_1} \right)}&0\\
    0&0&1
    \end{array}} \right]
    \end{array}
    \end{equation}

    \item $R_{{S^ * }}^S$ is the coordinate transformation matrix from the $\mathcal{S}^ * $ frame to the $\mathcal{S}$ frame, which can be written as
    \begin{equation}\label{eq:S*2S}
    R_{S^*}^S = \left[ {\begin{array}{*{20}{c}}
    1&{{\varphi_S}}&{ - {\eta_S}}\\
    { - {\varphi_S}}&1&{{\theta_S}}\\
    {{\eta_S}}&{ - {\theta_S}}&1
    \end{array}} \right]
    \end{equation}
\end{itemize}

Combining the Eqs.~\eqref{eq:OA*2OA}, \eqref{eq:OA2S}, \eqref{eq:OA*2S*}, and \eqref{eq:S*2S} into Eq.~\eqref{eq:transform}, similar steps would apply identically for the $\mathcal{OA}_2^*$ reference frame. Consequently, the resultant relationship is expressed as
\begin{equation}\label{eq:jitter2DWSFull}
\left\{ \begin{array}{l}
{\theta}_1^\text{DWS} = \cos(\psi_1){\theta}_\text{S} + \sin(\psi_1){\eta}_\text{S}, \\[1ex]
\eta_1^\text{DWS} =  \cos({\psi_{1}})\,{\eta_\text{S}} - \sin({\psi_{1}})\,{\theta_\text{S}}\\[1ex]
\varphi_1^\text{DWS} = {\varphi_\text{S}} - \Delta{\psi_1}\\[1ex]
{\theta}_2^\text{DWS} = \cos(\psi_2){\theta}_\text{S} + \sin(\psi_2){\eta}_\text{S}, \\[1ex]
\eta_2^\text{DWS} =  \cos({\psi_{2}})\,{\eta_\text{S}} - \sin({\psi_{2}})\,{\theta_\text{S}}\\[1ex]
\varphi_2^\text{DWS} = {\varphi_\text{S}} - \Delta{\psi_2}
\end{array} \right.
\end{equation}

\section{TTL noise in different TDI channel}\label{sec: AppendixB}
When utilizing alternative null channel combinations such as $C_3^{14}$, $C_{26}^{16}$, or $C_{27}^{16}$, we find that the time delays between the two signals on the same satellite exhibit slight differences. This subtle distinction enhances signal resolvability under correlated jitter conditions, reduces the condition number of the estimation matrix of least squares estimation, and ultimately improves estimation accuracy. Thus, in this appendix, we will clarify the derivation of how TTL noise is expressed within the noise equations for various null TDI channel combinations. Based on the Eq.~\eqref{eq:eta_k^TTL}, we write down all the terms of the jitter for the $\eta_{k}^{\text{TTL}}$ in $\eta_i$,
\begin{align}
    \eta_{k}^{\text{TTL}} &= C_{i,\eta}^{\text{RX}}{\eta}_i^\text{DWS} + C_{i,\varphi}^{\text{RX}}{\varphi}_i^\text{DWS} \nonumber\\
    &\quad + {\bf{D}}_{k}C_{j',\eta}^{\text{TX}}{\eta}_{j'}^\text{DWS} + {\bf{D}}_{k}C_{j',\varphi}^{\text{TX}}{\varphi}_{j'}^\text{DWS}.
\end{align}
For the sake of simplicity and readability, the unit conversion coefficients have been omitted. For the $\eta_{j'}^{\text{TTL}}$ in $\eta_{i'}$, we have 
\begin{align}
    \eta_{j'}^{\text{TTL}} &= C_{i',\eta}^{\text{RX}}{\eta}_{i'}^\text{DWS} + C_{i',\varphi}^{\text{RX}}{\varphi}_{i'}^\text{DWS} \nonumber\\
    &\quad + {\bf{D}}_{j'}C_{k,\eta}^{\text{TX}}{\eta}_{k}^\text{DWS} + {\bf{D}}_{j'}C_{k,\varphi}^{\text{TX}}{\varphi}_{k}^\text{DWS}.
\end{align}
By cyclic permutation of the indices, one can compute the remaining four $\eta^\text{TTL}$-terms in the $\eta$-terms, i.e., $\eta_{i}^{\text{TTL}}$ in $\eta_j$, $\eta_{k'}^{\text{TTL}}$ in $\eta_{j'}$, $\eta_{j}^{\text{TTL}}$ in $\eta_k$, $\eta_{i'}^{\text{TTL}}$ in $\eta_{k'}$.

In order to see the noise terms in the TDI combination, we are adding all the $\eta^\text{TTL}$-terms into the null TDI combination, for example, $C_3^{12}$, $C_3^{14}$ (two kinds of $\zeta$ channel), and $C_1^{16}$ (also known as X combination), as follows~\cite{hartwig2022characterization}
\begin{align}\label{eq:zeta_C312}
C_3^{12} & = (\bf{D}_{i'}\bf{D}_{i}\bf{A}_{j'} - \bf{D}_{j}\bf{A}_{k'}\bf{D}_{i}\bf{A}_{j'})\eta_{i'} \nonumber\\
&- (\bf{D}_{j}\bf{A}_{k'} - \bf{D}_{j}\bf{A}_{k'}\bf{D}_{i}\bf{A}_{j'}\bf{D}_{k})\eta_{j'} \nonumber\\
&+ (\bf{D}_{j}\bf{A}_{k'}\bf{D}_{i}\bf{A}_{j'} - \bf{D}_{i'}\bf{D}_{i}\bf{A}_{j'})\eta_{i} \nonumber\\
&- (\bf{D}_{i'} - \bf{D}_{j}\bf{A}_{k'})\eta_{j}  \nonumber\\
&+ (1 - \bf{D}_{i'}\bf{D}_{i}\bf{A}_{j'}\bf{D}_{k}\bf{A}_{i'})\eta_{k} \nonumber\\
&- (1 - \bf{D}_{i'}\bf{D}_{i}\bf{A}_{j'}\bf{D}_{k}\bf{A}_{i'})\eta_{k'}~,
\end{align}
\begin{align}\label{eq:zeta_C314}
C_3^{14} &= (\bf{D}_{i}\bf{A}_{j'}-\bf{D}_{k'}\bf{D}_{k}\bf{D}_{i}\bf{A}_{j'} + \bf{D}_{i}\bf{A}_{j'}\bf{D}_{k}\bf{A}_{i'}\bf{D}_{j} - \bf{D}_{k'})\eta_{i} \nonumber\\
&- (\bf{D}_{i}\bf{A}_{j'} - \bf{D}_{k'}\bf{D}_{k}\bf{D}_{i}\bf{A}_{j'})\eta_{i'} \nonumber\\
&- (\bf{D}_{i}\bf{A}_{j'}\bf{D}_{k}\bf{A}_{i'} - \bf{D}_{k'}\bf{D}_{k}\bf{D}_{i}\bf{A}_{j'}\bf{D}_{k}\bf{A}_{i'})\eta_{k'} \nonumber\\
&- (1 - \bf{D}_{i}\bf{A}_{j'}\bf{D}_{k}\bf{A}_{i'}\bf{D}_{j}\bf{D}_{k})\eta_{j'} \nonumber\\
&+ (\bf{D}_{i}\bf{A}_{j'}\bf{D}_{k}\bf{A}_{i'} - \bf{D}_{k'}\bf{D}_{k}\bf{D}_{i}A\bf{A}_{j'}\bf{D}_{k}\bf{A}_{i'})\eta_{k} \nonumber\\
&+ (1 - \bf{D}_{k'}\bf{D}_{k})\eta_{j}~,
\end{align}
and 
\begin{align} \label{eq:X}
C_1^{16} &= (1 - \bf{D}_{j'}\bf{D}_{j} - \bf{D}_{j'}\bf{D}_{j}\bf{D}_{k}\bf{D}_{k'} + \bf{D}_{k}\bf{D}_{k'}\bf{D}_{j'}\bf{D}_{j}\bf{D}_{j'}\bf{D}_{j})\eta_{i}\nonumber\\
&- (1 - \bf{D}_{k}\bf{D}_{k'} - \bf{D}_{k}\bf{D}_{k'}\bf{D}_{j'}\bf{D}_{j} + \bf{D}_{j'}\bf{D}_{j}\bf{D}_{k}\bf{D}_{k'}\bf{D}_{k}\bf{D}_{k'})\eta_{i'} \nonumber\\
&+ (\bf{D}_{k} - \bf{D}_{j'}\bf{D}_{j}\bf{D}_{k} - \bf{D}_{j'}\bf{D}_{j}\bf{D}_{k}\bf{D}_{k'}\bf{D}_{k} \nonumber\\
&+ \bf{D}_{k}\bf{D}_{k'}\bf{D}_{j'}\bf{D}_{j}\bf{D}_{j'}\bf{D}_{j}\bf{D}_{k})\eta_{j'} \nonumber\\
&- (\bf{D}_{j'} - \bf{D}_{k}\bf{D}_{k'}\bf{D}_{j'}\bf{D}_{j}\bf{D}_{j'} \nonumber \\
&+ \bf{D}_{j'}\bf{D}_{j}\bf{D}_{k}\bf{D}_{k'}\bf{D}_{k}\bf{D}_{k'}\bf{D}_{j'} - \bf{D}_{k}\bf{D}_{k'}\bf{D}_{j'})\eta_{k}~.
\end{align}

The following expressions represent the TTL noise terms under these specific combination conditions. Due to the symmetry of TDI null channel combinations, the order of delay and advance symbols correspond to each other. For conciseness, we categorize these two symbols separately to better demonstrate similar time delays. For $C_3^{12}$,
\begin{widetext}\label{eq:C312Full}
\begin{align}
C_3^{12} &= -(\bf{A}_{k'}\bf{A}_{j'})\left(\bf{D}_{i}\bf{D}_{j}C_{i,\eta}^{\text{RX}}{\eta}_{i}^\text{DWS} - \bf{D}_{i}\bf{D}_{j}C_{i',\eta}^{\text{RX}}{\eta}_{i'}^\text{DWS} + \bf{D}_{j}\bf{D}_{j'}C_{j,\eta}^{\text{RX}}{\eta}_{j}^\text{DWS} - \bf{D}_{j}\bf{D}_{j'}C_{j',\eta}^{\text{RX}}{\eta}_{j'}^\text{DWS} + \bf{D}_{j'}\bf{D}_{k'}C_{k,\eta}^{\text{RX}}{\eta}_{k}^\text{DWS} \right. \nonumber\\
&\left. - \bf{D}_{j'}\bf{D}_{k'}C_{k',\eta}^{\text{RX}}{\eta}_{k'}^\text{DWS} + \bf{D}_{i}\bf{D}_{j}C_{i,\varphi}^{\text{RX}}{\varphi}_{i}^\text{DWS} - \bf{D}_{i}\bf{D}_{j}C_{i',\varphi}^{\text{RX}}{\varphi}_{i'}^\text{DWS} + \bf{D}_{j}\bf{D}_{j'}C_{j,\varphi}^{\text{RX}}{\varphi}_{j}^\text{DWS} - \bf{D}_{j}\bf{D}_{j'}C_{j',\varphi}^{\text{RX}}{\varphi}_{j'}^\text{DWS} \right. \nonumber\\
&\left. + \bf{D}_{j'}\bf{D}_{k'}C_{k,\varphi}^{\text{RX}}{\varphi}_{k}^\text{DWS} - \bf{D}_{j'}\bf{D}_{k'}C_{k',\varphi}^{\text{RX}}{\varphi}_{k'}^\text{DWS} - \bf{D}_{i}\bf{D}_{i'}\bf{D}_{k'}C_{i,\varphi}^{\text{RX}}{\varphi}_{i}^\text{DWS} + \bf{D}_{i}\bf{D}_{i'}\bf{D}_{k'}C_{i',\varphi}^{\text{RX}}{\varphi}_{i'}^\text{DWS} + \bf{D}_{i}\bf{D}_{j}\bf{D}_{k}C_{j',\varphi}^{\text{RX}}{\varphi}_{j'}^\text{DWS} \right. \nonumber\\
&\left. - \bf{D}_{i'}\bf{D}_{j'}\bf{D}_{k'}C_{j,\varphi}^{\text{RX}}{\varphi}_{j}^\text{DWS} + \bf{D}_{j}\bf{D}_{j'}\bf{D}_{k'}C_{i,\varphi}^{\text{TX}}{\varphi}_{i}^\text{DWS} - \bf{D}_{j}\bf{D}_{j'}\bf{D}_{k'}C_{i',\varphi}^{\text{TX}}{\varphi}_{i'}^\text{DWS} - \bf{D}_{i}\bf{D}_{j}\bf{D}_{k}C_{j',\varphi}^{\text{TX}}{\varphi}_{j'}^\text{DWS} + \bf{D}_{i'}\bf{D}_{j'}\bf{D}_{k'}C_{j,\varphi}^{\text{TX}}{\varphi}_{j}^\text{DWS} \right. \nonumber\\
&\left. + \bf{D}_{i}\bf{D}_{j}\bf{D}_{j'}C_{k,\varphi}^{\text{TX}}{\varphi}_{k}^\text{DWS} - \bf{D}_{i}\bf{D}_{j}\bf{D}_{j'}C_{k',\varphi}^{\text{TX}}{\varphi}_{k'}^\text{DWS} - \bf{D}_{i}\bf{D}_{k}\bf{D}_{k'}C_{k,\varphi}^{\text{RX}}{\varphi}_{k}^\text{DWS} + \bf{D}_{i}\bf{D}_{k}\bf{D}_{k'}C_{k',\varphi}^{\text{RX}}{\varphi}_{k'}^\text{DWS} - \bf{D}_{i}\bf{D}_{i'}\bf{D}_{k'}C_{i,\eta}^{\text{RX}}{\eta}_{i}^\text{DWS} \right. \nonumber\\
&\left. + \bf{D}_{i}\bf{D}_{i'}\bf{D}_{k'}C_{i',\eta}^{\text{RX}}{\eta}_{i'}^\text{DWS} + \bf{D}_{i}\bf{D}_{j}\bf{D}_{k}C_{j',\eta}^{\text{RX}}{\eta}_{j'}^\text{DWS} - \bf{D}_{i'}\bf{D}_{j'}\bf{D}_{k'}C_{j,\eta}^{\text{RX}}{\eta}_{j}^\text{DWS} + \bf{D}_{j}\bf{D}_{j'}\bf{D}_{k'}C_{i,\eta}^{\text{TX}}{\eta}_{i}^\text{DWS} - \bf{D}_{j}\bf{D}_{j'}\bf{D}_{k'}C_{i',\eta}^{\text{TX}}{\eta}_{i'}^\text{DWS} \right. \nonumber\\
&\left. - \bf{D}_{i}\bf{D}_{j}\bf{D}_{k}C_{j',\eta}^{\text{TX}}{\eta}_{j'}^\text{DWS} + \bf{D}_{i'}\bf{D}_{j'}\bf{D}_{k'}C_{j,\eta}^{\text{TX}}{\eta}_{j}^\text{DWS} + \bf{D}_{i}\bf{D}_{j}\bf{D}_{j'}C_{k,\eta}^{\text{TX}}{\eta}_{k}^\text{DWS} - \bf{D}_{i}\bf{D}_{j}\bf{D}_{j'}C_{k',\eta}^{\text{TX}}{\eta}_{k'}^\text{DWS} - \bf{D}_{i}\bf{D}_{k}\bf{D}_{k'}C_{k,\eta}^{\text{RX}}{\eta}_{k}^\text{DWS} \right. \nonumber\\
&\left. + \bf{D}_{i}\bf{D}_{k}\bf{D}_{k'}C_{k',\eta}^{\text{RX}}{\eta}_{k'}^\text{DWS} - \bf{D}_{i}\bf{D}_{j}\bf{D}_{k}\bf{D}_{k'}C_{i,\eta}^{\text{TX}}{\eta}_{i}^\text{DWS} + \bf{D}_{i}\bf{D}_{j}\bf{D}_{k}\bf{D}_{k'}C_{i',\eta}^{\text{TX}}{\eta}_{i'}^\text{DWS} - \bf{D}_{i}\bf{D}_{i'}\bf{D}_{k}\bf{D}_{k'}C_{j,\eta}^{\text{TX}}{\eta}_{j}^\text{DWS} \right. \nonumber\\
&\left. + \bf{D}_{i}\bf{D}_{i'}\bf{D}_{k}\bf{D}_{k'}C_{j',\eta}^{\text{TX}}{\eta}_{j'}^\text{DWS} - \bf{D}_{i}\bf{D}_{i'}\bf{D}_{j'}\bf{D}_{k'}C_{k,\eta}^{\text{TX}}{\eta}_{k}^\text{DWS} + \bf{D}_{i}\bf{D}_{i'}\bf{D}_{j'}\bf{D}_{k'}C_{k',\eta}^{\text{TX}}{\eta}_{k'}^\text{DWS} - \bf{D}_{i}\bf{D}_{j}\bf{D}_{k}\bf{D}_{k'}C_{i,\varphi}^{\text{TX}}{\varphi}_{i}^\text{DWS} \right. \nonumber\\
&\left.  + \bf{D}_{i}\bf{D}_{j}\bf{D}_{k}\bf{D}_{k'}C_{i',\varphi}^{\text{TX}}{\varphi}_{i'}^\text{DWS} - \bf{D}_{i}\bf{D}_{i'}\bf{D}_{k}\bf{D}_{k'}C_{j,\varphi}^{\text{TX}}{\varphi}_{j}^\text{DWS} + \bf{D}_{i}\bf{D}_{i'}\bf{D}_{k}\bf{D}_{k'}C_{j',\varphi}^{\text{TX}}{\varphi}_{j'}^\text{DWS} \right. \nonumber\\
&\left. - \bf{D}_{i}\bf{D}_{i'}\bf{D}_{j'}\bf{D}_{k'}C_{k,\varphi}^{\text{TX}}{\varphi}_{k}^\text{DWS} + \bf{D}_{i}\bf{D}_{i'}\bf{D}_{j'}\bf{D}_{k'}C_{k',\varphi}^{\text{TX}}{\varphi}_{k'}^\text{DWS}\right).
\end{align}
\end{widetext}
The equation demonstrates that pitch and yaw jitter noises for each pair $i$ and $i'$, $j$ and $j'$, $k$ and $k'$ share an identical time delay configuration. Consequently, under the correlated jitter scenario for a specific pair of MOSAs, distinguishing the coefficient becomes challenging. Consequently, while certain coefficients can be estimated to account for the TTL noise subtraction within the null channel, these estimated coefficients are not valid for other combinations, such as X. 

For the $C_3^{14}$ combination, we have
\begin{widetext}\label{eq:C314Full}
\begin{align}
C_3^{14} &= -\left(\bf{A}_{j'}\bf{A}_{i'}\right)\left(\bf{D}_{i}\bf{D}_{i'}C_{i,\eta}^{\text{RX}}{\eta}_{i}^\text{DWS} - \bf{D}_{i}\bf{D}_{i'}C_{i',\eta}^{\text{RX}}{\eta}_{i'}^\text{DWS} + \bf{D}_{i'}\bf{D}_{j'}C_{j,\eta}^{\text{RX}}{\eta}_{j}^\text{DWS} - \bf{D}_{i'}\bf{D}_{j'}C_{j',\eta}^{\text{RX}}{\eta}_{j'}^\text{DWS} + \bf{D}_{i}\bf{D}_{k}C_{k,\eta}^{\text{RX}}{\eta}_{k}^\text{DWS} \right. \nonumber\\
&\left. - \bf{D}_{i}\bf{D}_{k}C_{k',\eta}^{\text{RX}}{\eta}_{k'}^\text{DWS} + \bf{D}_{i}\bf{D}_{i'}C_{i,\varphi}^{\text{RX}}{\varphi}_{i}^\text{DWS} - \bf{D}_{i}\bf{D}_{i'}C_{i',\varphi}^{\text{RX}}{\varphi}_{i'}^\text{DWS} + \bf{D}_{i'}\bf{D}_{j'}C_{j,\varphi}^{\text{RX}}{\varphi}_{j}^\text{DWS} - \bf{D}_{i'}\bf{D}_{j'}C_{j',\varphi}^{\text{RX}}{\varphi}_{j'}^\text{DWS} \right. \nonumber\\
&\left. + \bf{D}_{i}\bf{D}_{k}C_{k,\varphi}^{\text{RX}}{\varphi}_{k}^\text{DWS} - \bf{D}_{i}\bf{D}_{k}C_{k',\varphi}^{\text{RX}}{\varphi}_{k'}^\text{DWS} + \bf{D}_{i}\bf{D}_{j}\bf{D}_{k}C_{i,\varphi}^{\text{RX}}{\varphi}_{i}^\text{DWS} - \bf{D}_{i'}\bf{D}_{j'}\bf{D}_{k'}C_{i,\varphi}^{\text{RX}}{\varphi}_{i}^\text{DWS} - \bf{D}_{i}\bf{D}_{j}\bf{D}_{k}C_{i',\varphi}^{\text{TX}}{\varphi}_{i'}^\text{DWS} \right. \nonumber\\
&\left. + \bf{D}_{i'}\bf{D}_{j'}\bf{D}_{k'}C_{i,\varphi}^{\text{TX}}{\varphi}_{i}^\text{DWS} + \bf{D}_{i}\bf{D}_{i'}\bf{D}_{k}C_{j,\varphi}^{\text{TX}}{\varphi}_{j}^\text{DWS} - \bf{D}_{i}\bf{D}_{i'}\bf{D}_{k}C_{j',\varphi}^{\text{TX}}{\varphi}_{j'}^\text{DWS} + \bf{D}_{i}\bf{D}_{i'}\bf{D}_{j'}C_{k,\varphi}^{\text{TX}}{\varphi}_{k}^\text{DWS} - \bf{D}_{i}\bf{D}_{i'}\bf{D}_{j'}C_{k',\varphi}^{\text{TX}}{\varphi}_{k'}^\text{DWS} \right. \nonumber\\
&\left. + \bf{D}_{i}\bf{D}_{j}\bf{D}_{k}^2C_{j',\eta}^{\text{RX}}{\eta}_{j'}^\text{DWS} - \bf{D}_{i}\bf{D}_{j}\bf{D}_{k}^2C_{j',\eta}^{\text{TX}}{\eta}_{j'}^\text{DWS} - \bf{D}_{i}\bf{D}_{k}^2\bf{D}_{k'}C_{k,\eta}^{\text{RX}}{\eta}_{k}^\text{DWS} + \bf{D}_{i}\bf{D}_{k}^2\bf{D}_{k'}C_{k',\eta}^{\text{RX}}{\eta}_{k'}^\text{DWS} + \bf{D}_{i}\bf{D}_{j}\bf{D}_{k}^2C_{j',\varphi}^{\text{RX}}{\varphi}_{j'}^\text{DWS} \right. \nonumber\\
&\left. - \bf{D}_{i}\bf{D}_{j}\bf{D}_{k}^2C_{j',\varphi}^{\text{TX}}{\varphi}_{j'}^\text{DWS} - \bf{D}_{i}\bf{D}_{k}^2\bf{D}_{k'}C_{k,\varphi}^{\text{RX}}{\varphi}_{k}^\text{DWS} + \bf{D}_{i}\bf{D}_{k}^2\bf{D}_{k'}C_{k',\varphi}^{\text{RX}}{\varphi}_{k'}^\text{DWS} + \bf{D}_{i}\bf{D}_{j}\bf{D}_{k}C_{i,\eta}^{\text{RX}}{\eta}_{i}^\text{DWS} - \bf{D}_{i'}\bf{D}_{j'}\bf{D}_{k'}C_{i,\eta}^{\text{RX}}{\eta}_{i}^\text{DWS} \right. \nonumber\\
&\left. - \bf{D}_{i}\bf{D}_{j}\bf{D}_{k}C_{i',\eta}^{\text{TX}}{\eta}_{i'}^\text{DWS} + \bf{D}_{i'}\bf{D}_{j'}\bf{D}_{k'}C_{i,\eta}^{\text{TX}}{\eta}_{i}^\text{DWS} + \bf{D}_{i}\bf{D}_{i'}\bf{D}_{k}C_{j,\eta}^{\text{TX}}{\eta}_{j}^\text{DWS} - \bf{D}_{i}\bf{D}_{i'}\bf{D}_{k}C_{j',\eta}^{\text{TX}}{\eta}_{j'}^\text{DWS} + \bf{D}_{i}\bf{D}_{i'}\bf{D}_{j'}C_{k,\eta}^{\text{TX}}{\eta}_{k}^\text{DWS}\right. \nonumber\\
&\left. - \bf{D}_{i}\bf{D}_{i'}\bf{D}_{j'}C_{k',\eta}^{\text{TX}}{\eta}_{k'}^\text{DWS} - \bf{D}_{i}\bf{D}_{i'}\bf{D}_{k}\bf{D}_{k'}C_{i,\eta}^{\text{RX}}{\eta}_{i}^\text{DWS} + \bf{D}_{i}\bf{D}_{i'}\bf{D}_{k}\bf{D}_{k'}C_{i',\eta}^{\text{RX}}{\eta}_{i'}^\text{DWS} - \bf{D}_{i'}\bf{D}_{j'}\bf{D}_{k}\bf{D}_{k'}C_{j,\eta}^{\text{RX}}{\eta}_{j}^\text{DWS} \right. \nonumber\\
&\left. + \bf{D}_{i'}\bf{D}_{j'}\bf{D}_{k}\bf{D}_{k'}C_{j',\eta}^{\text{TX}}{\eta}_{j'}^\text{DWS} - \bf{D}_{i}\bf{D}_{i'}\bf{D}_{k}\bf{D}_{k'}C_{i,\varphi}^{\text{RX}}{\varphi}_{i}^\text{DWS} + \bf{D}_{i}\bf{D}_{i'}\bf{D}_{k}\bf{D}_{k'}C_{i',\varphi}^{\text{RX}}{\varphi}_{i'}^\text{DWS} - \bf{D}_{i'}\bf{D}_{j'}\bf{D}_{k}\bf{D}_{k'}C_{j,\varphi}^{\text{RX}}{\varphi}_{j}^\text{DWS} \right. \nonumber\\
&\left. + \bf{D}_{i'}\bf{D}_{j'}\bf{D}_{k}\bf{D}_{k'}C_{j',\varphi}^{\text{TX}}{\varphi}_{j'}^\text{DWS} - \bf{D}_{i}\bf{D}_{j}\bf{D}_{k}^2\bf{D}_{k'}C_{i,\eta}^{\text{TX}}{\eta}_{i}^\text{DWS} + \bf{D}_{i}\bf{D}_{j}\bf{D}_{k}^2\bf{D}_{k'}C_{i',\eta}^{\text{TX}}{\eta}_{i'}^\text{DWS} - \bf{D}_{i}\bf{D}_{i'}\bf{D}_{k}^2\bf{D}_{k'}C_{j,\eta}^{\text{TX}}{\eta}_{j}^\text{DWS} \right. \nonumber\\
&\left. + \bf{D}_{i}\bf{D}_{i'}\bf{D}_{k}^2\bf{D}_{k'}C_{j',\eta}^{\text{TX}}{\eta}_{j'}^\text{DWS} - \bf{D}_{i}\bf{D}_{j}\bf{D}_{k}^2\bf{D}_{k'}C_{i,\varphi}^{\text{TX}}{\varphi}_{i}^\text{DWS} + \bf{D}_{i}\bf{D}_{j}\bf{D}_{k}^2\bf{D}_{k'}C_{i',\varphi}^{\text{TX}}{\varphi}_{i'}^\text{DWS} - \bf{D}_{i}\bf{D}_{i'}\bf{D}_{k}^2\bf{D}_{k'}C_{j,\varphi}^{\text{TX}}{\varphi}_{j}^\text{DWS} \right. \nonumber\\
&\left. + \bf{D}_{i}\bf{D}_{i'}\bf{D}_{k}^2\bf{D}_{k'}C_{j',\varphi}^{\text{TX}}{\varphi}_{j'}^\text{DWS} - \bf{D}_{i}\bf{D}_{i'}\bf{D}_{j'}\bf{D}_{k}\bf{D}_{k'}C_{k,\eta}^{\text{TX}}{\eta}_{k}^\text{DWS} + \bf{D}_{i}\bf{D}_{i'}\bf{D}_{j'}\bf{D}_{k}\bf{D}_{k'}C_{k',\eta}^{\text{TX}}{\eta}_{k'}^\text{DWS} - \bf{D}_{i}\bf{D}_{i'}\bf{D}_{j'}\bf{D}_{k}\bf{D}_{k'}C_{k,\varphi}^{\text{TX}}{\varphi}_{k}^\text{DWS} \right. \nonumber\\
&\left. + \bf{D}_{i}\bf{D}_{i'}\bf{D}_{j'}\bf{D}_{k}\bf{D}_{k'}C_{k',\varphi}^{\text{TX}}{\varphi}_{k'}^\text{DWS}\right).
\end{align}
\end{widetext}

As seen in the preceding equation, the presence of varied time delay groups for certain terms marginally intensifies the discrepancies between signals, e.g.
\begin{widetext}
$C_{i,\varphi}^{\text{RX}}{\varphi}_{i}^\text{DWS}$, $C_{i,\varphi}^{\text{RX}}{\varphi}_{i}^\text{DWS}$, $C_{i',\varphi}^{\text{TX}}{\varphi}_{i'}^\text{DWS}$, $C_{i,\varphi}^{\text{TX}}{\varphi}_{i}^\text{DWS}$, $C_{j',\eta}^{\text{RX}}{\eta}_{j'}^\text{DWS}$, $C_{j',\eta}^{\text{TX}}{\eta}_{j'}^\text{DWS}$, $C_{j',\varphi}^{\text{RX}}{\varphi}_{j'}^\text{DWS}$, $C_{j',\varphi}^{\text{TX}}{\varphi}_{j'}^\text{DWS}$, $C_{i,\eta}^{\text{RX}}{\eta}_{i}^\text{DWS}$, $ C_{i,\eta}^{\text{RX}}{\eta}_{i}^\text{DWS}$, $ C_{i',\eta}^{\text{TX}}{\eta}_{i'}^\text{DWS}$, and $ C_{i,\eta}^{\text{TX}}{\eta}_{i}^\text{DWS}$,
\end{widetext}
thus improving estimation accuracy.

To consider the Fig.\ref{fig:C312C314Demo}, in $C_3^{14}$, it can be observed that there are six links between $i$ and $j'$, while there are four links each between $i'$ and $k$, and between $j$ and $k'$. In contrast, the $C_3^{12}$ configuration provides an equal number of four paths for each satellite pair, which enhances symmetry but diminishes the effectiveness of the estimation process.

For $C_1^{16}$, 
\begin{widetext}\label{eq:C116Full}
\begin{align}
C_1^{16} &= (\bf{D}_{j}\bf{D}_{j'}\bf{D}_{k}\bf{D}_{k'} - 1)\left(C_{i,\eta}^{\text{RX}}{\eta}_{i}^\text{DWS} - C_{i',\eta}^{\text{RX}}{\eta}_{i'}^\text{DWS} + C_{i,\varphi}^{\text{RX}}{\varphi}_{i}^\text{DWS} - C_{i',\varphi}^{\text{RX}}{\varphi}_{i'}^\text{DWS} + \bf{D}_{k}C_{j',\eta}^{\text{RX}}{\eta}_{j'}^\text{DWS} - \bf{D}_{j'}C_{k,\eta}^{\text{RX}}{\eta}_{k}^\text{DWS}\right. \nonumber\\
&\left. - \bf{D}_{k}C_{j',\eta}^{\text{TX}}{\eta}_{j'}^\text{DWS} + \bf{D}_{j'}C_{k,\eta}^{\text{TX}}{\eta}_{k}^\text{DWS} + \bf{D}_{k}C_{j',\varphi}^{\text{RX}}{\varphi}_{j'}^\text{DWS} - \bf{D}_{j'}C_{k,\varphi}^{\text{RX}}{\varphi}_{k}^\text{DWS} - \bf{D}_{k}C_{j',\varphi}^{\text{TX}}{\varphi}_{j'}^\text{DWS} + \bf{D}_{j'}C_{k,\varphi}^{\text{TX}}{\varphi}_{k}^\text{DWS} \right. \nonumber\\
&\left. - \bf{D}_{j}\bf{D}_{j'}C_{i,\eta}^{\text{RX}}{\eta}_{i}^\text{DWS} + \bf{D}_{j}\bf{D}_{j'}C_{i',\eta}^{\text{TX}}{\eta}_{i'}^\text{DWS} + \bf{D}_{k}\bf{D}_{k'}C_{i',\eta}^{\text{RX}}{\eta}_{i'}^\text{DWS} - \bf{D}_{k}\bf{D}_{k'}C_{i,\eta}^{\text{TX}}{\eta}_{i}^\text{DWS} - \bf{D}_{j}\bf{D}_{j'}C_{i,\varphi}^{\text{RX}}{\varphi}_{i}^\text{DWS} + \bf{D}_{j}\bf{D}_{j'}C_{i',\varphi}^{\text{TX}}{\varphi}_{i'}^\text{DWS} \right. \nonumber\\
&\left. + \bf{D}_{k}\bf{D}_{k'}C_{i',\varphi}^{\text{RX}}{\varphi}_{i'}^\text{DWS} - \bf{D}_{k}\bf{D}_{k'}C_{i,\varphi}^{\text{TX}}{\varphi}_{i}^\text{DWS} - \bf{D}_{j'}\bf{D}_{k}\bf{D}_{k'}C_{k,\eta}^{\text{TX}}{\eta}_{k}^\text{DWS} - \bf{D}_{j}\bf{D}_{j'}\bf{D}_{k}C_{j',\varphi}^{\text{RX}}{\varphi}_{j'}^\text{DWS} + \bf{D}_{j}\bf{D}_{j'}\bf{D}_{k}C_{j',\varphi}^{\text{TX}}{\varphi}_{j'}^\text{DWS} \right. \nonumber\\
&\left. + \bf{D}_{j'}\bf{D}_{k}\bf{D}_{k'}C_{k,\varphi}^{\text{RX}}{\varphi}_{k}^\text{DWS} - \bf{D}_{j'}\bf{D}_{k}\bf{D}_{k'}C_{k,\varphi}^{\text{TX}}{\varphi}_{k}^\text{DWS} - \bf{D}_{j}\bf{D}_{j'}\bf{D}_{k}C_{j',\eta}^{\text{RX}}{\eta}_{j'}^\text{DWS} + \bf{D}_{j}\bf{D}_{j'}\bf{D}_{k}C_{j',\eta}^{\text{TX}}{\eta}_{j'}^\text{DWS} + \bf{D}_{j'}\bf{D}_{k}\bf{D}_{k'}C_{k,\eta}^{\text{RX}}{\eta}_{k}^\text{DWS} \right. \nonumber\\
&\left. + \bf{D}_{j}\bf{D}_{j'}\bf{D}_{k}\bf{D}_{k'}C_{i,\eta}^{\text{TX}}{\eta}_{i}^\text{DWS} - \bf{D}_{j}\bf{D}_{j'}\bf{D}_{k}\bf{D}_{k'}C_{i',\eta}^{\text{TX}}{\eta}_{i'}^\text{DWS} + \bf{D}_{j}\bf{D}_{j'}\bf{D}_{k}\bf{D}_{k'}C_{i,\varphi}^{\text{TX}}{\varphi}_{i}^\text{DWS} - \bf{D}_{j}\bf{D}_{j'}\bf{D}_{k}\bf{D}_{k'}C_{i',\varphi}^{\text{TX}}{\varphi}_{i'}^\text{DWS}\right).
\end{align}
\end{widetext}
In this scenario, if the two yaw jitters on the same satellite are similar, $i$ and $i'$ become indistinguishable, whereas $j'$ and $k$ remain distinguishable. Therefore, employing the X, Y, Z channel (i.e., the X combination with three satellites 1, 2, 3) effectively addresses the issues of jitter correlation.

Through this analysis, it has been determined that to enhance the estimation of coefficients, one may either amplify the distinctions within the jitter noise, or employ various TDI combinations to change the sequence of time delays for jitter signals $i$ and $i'$ on each satellite. Building on these insights, the simulation results recommend the use of high-frequency data and a single MOSA rotation strategy for coefficient estimation in Sec.\ref{sec: result}. Furthermore, the discussion section \ref{sec: discussion} explores the feasibility and validation of the adoption of the $C_3^{14}$ combination and the modulation of MOSA jitter.


\nocite{*}

\bibliography{bibliography}
\end{document}